\documentclass[11pt]{article}
\usepackage{amsmath}
\usepackage{amssymb}
\usepackage{amsfonts,amstext,latexsym,xspace}
\usepackage{epsfig}

\makeatletter
\newif\if@fewtab\@fewtabtrue
\def\moth{\mathsurround=0pt}
\newdimen\zo \zo=0pt

\def\tick{\leaders\hrule height 0.5ex depth 0pt \hskip 0.5pt}
\def\upboxfill{$\moth \setbox\zo\hbox{\tick}%
  \hskip 2pt\hbox to 0pt{$\tick$\hss}\hrulefill \hbox to 2pt{$\tick$\hss}$}

\def\dtick{\leaders\hrule height .34pt depth 0.5ex \hskip 0.5pt}
\def\downboxfill{$\moth \setbox\zo\hbox{\dtick}%
  \hskip 2pt\hbox to 0pt{$\dtick$\hss}\hrulefill%
  \hbox to 2pt{$\dtick$\hss}$}

\newcommand{\cX}{\mathcal{X}}

\newcommand{\Sympl}[2]{\left\langle #1,#2\right\rangle}
\newcommand{\FTP}[4][{}]{\left( #2,#3,#4 \right)^{#1}}

\newcommand{\Cn}{\ensuremath{\mathbb{C}}\xspace}

\begin{document}

\begin{titlepage}
\begin{center}
\begin{large}
\textbf{Generalized spacetimes defined by cubic forms and the
minimal unitary realizations of their quasiconformal groups}
\end{large}
\vspace{1cm}
\\

\begin{large}
Murat G\"{u}naydin$^{\dagger}$\footnote{murat@phys.psu.edu} and
Oleksandr Pavlyk$^{\dagger, \ddagger}$\footnote{pavlyk@phys.psu.edu}
\end{large}
\\
\vspace{.35cm}
$^{\dagger}$ \emph{Physics Department \\
Pennsylvania State University\\
University Park, PA 16802, USA} \\
\vspace{.3cm}
and \\
\vspace{.3cm}
$^{\ddagger}$
\emph{Wolfram Research Inc. \\100 Trade Center Dr. \\ Champaign, IL 61820, USA} \\

\begin{abstract}
{\small We study the symmetries of generalized spacetimes and
 corresponding phase spaces defined by Jordan algebras of degree
 three. The generic Jordan family of formally real Jordan algebras of
 degree three describe extensions of the Minkowskian spacetimes
 by an extra ``dilatonic'' coordinate, whose rotation, Lorentz and
 conformal groups are $\mathrm{SO}(d-1), \mathrm{SO}(d-1,1)\times
 \mathrm{SO}(1,1)$ and $\mathrm{SO}(d,2)\times \mathrm{SO}(2,1)$,
 respectively. The generalized spacetimes described by simple Jordan
 algebras of degree three correspond to extensions of Minkowskian
 spacetimes in the critical dimensions $(d=3,4,6,10)$ by a dilatonic
 and extra ($2,4,8,16$) commuting spinorial coordinates,
 respectively. Their rotation, Lorentz and conformal groups are those
 that occur in the first three rows of the Magic Square. The
 Freudenthal triple systems defined over these Jordan algebras
 describe conformally covariant phase spaces. Following
 hep-th/0008063, we give a unified geometric realization of the
 quasiconformal groups that act on their conformal phase spaces
 extended by an extra ``cocycle'' coordinate. For the generic Jordan
 family the quasiconformal groups are $\mathrm{SO}(d+2,4)$, whose minimal
 unitary realizations are given. The minimal unitary representations of
  the quasiconformal groups  $\mathrm{F}_{4(4)}$, $\mathrm{E}_{6(2)}$,
 $\mathrm{E}_{7(-5)}$ and $\mathrm{E}_{8(-24)}$ of the simple Jordan family
  were given in our earlier work hep-th/0409272. }
\end{abstract}
\end{center}
\end{titlepage}
\newpage
\section{Introduction}

One can define generalized space-times coordinatized by Jordan
algebras $J$, in such a way that their rotation, Lorentz, and
conformal groups can be identified with the automorphism, reduced
structure, and the linear fractional groups of of the corresponding
Jordan algebras $J$, respectively~\cite{mg75,mg80,mg91,mg92}. The main
requirements for formulating relativistic quantum field theories over
four dimensional Minkowski spacetime extend naturally to the
generalized space times defined by formally real Jordan algebras. For
example, the well-known connection between the positive energy unitary
representations of the four dimensional conformal group
$\mathrm{SU}(2,2)$ and the covariant fields transforming in finite
dimensional representations of the Lorentz group
$\mathrm{SL}(2,\Cn)$~\cite{Mack69,Mack77} extends to all generalized
space-times defined by formally real Jordan algebras~\cite{Guna99}.

Except for $\mathrm{G}_2, \mathrm{F}_4$ and $\mathrm{E}_8$, certain
noncompact real forms of all simple groups arise as conformal groups
of formally real Jordan algebras. The corresponding real forms are
precisely the ones that admit positive energy unitary
representations. These conformal groups act geometrically on the
corresponding Jordan algebras via fractional linear transformations
\cite{Koecher}. In particular, the exceptional group $\mathrm{E}_7$
acts as the conformal group of the exceptional Jordan
algebra\cite{mg75}. For the real exceptional Jordan algebra
$J_3^{\mathbb{O}}$ over the division algebra of real octonions the
conformal group is $\mathrm{E}_{7(-25)}$, which admits positive
energy unitary representations.

Motivated by possible applications to the U-duality groups of
maximal supergravity and M-theory, the first geometric realization
of $\mathrm{E}_{8(8)}$ was given in \cite{GKN1}. This geometric
realization of $\mathrm{E}_{8(8)}$ on a 57 dimensional space was
called quasiconformal since it leaves invariant a suitably defined
light cone with respect to a quartic norm. This construction of
$\mathrm{E}_{8(8)}$ together with the corresponding construction of
$\mathrm{E}_{8(-24)}$ \cite{GP}  contain all previous geometric
realizations of the symmetries of generalized space-times based on
exceptional Lie groups, and at the same time goes beyond the
framework of Jordan algebras.

The algebraic structures related to the novel quasiconformal
realizations are not Jordan algebras, but rather Freudenthal triple
systems (FTS)\cite{Freu54,Meyb68}. The 57 dimensional space on which
$\mathrm{E}_{8(8)}$ or $E_{8(-24)}$ is realized is the direct sum of
a 56 dimensional space of the underlying FTS and a singlet defined
by the symplectic invariant over that FTS. The nonlinear realization
of the Lie algebra of $\mathrm{E}_{8(8)}$ given in \cite{GKN1} was
written in terms of the triple product of the FTS and hence extends
to all simple Lie algebras since they all can be constructed over
FTS's. This follows from the fact that simple Freudenthal triple
systems are in one-to-one correspondence with simple Lie algebras
with a five graded decomposition
\begin{equation}
\mathfrak{g} = \mathfrak{g}^{-2} \oplus \mathfrak{g}^{-1} \oplus
    \mathfrak{g}^0 \oplus \mathfrak{g}^{+1}
      \oplus \mathfrak{g}^{+2} \,.  \label{5grading}
\end{equation}
such that grade $\pm 2$ spaces are one dimensional \cite{KanSko82}.
These include all simple Lie algebras, except for $\mathfrak{sl}(2)$
for which it degenerates to a 3-graded structure.

The formally real Jordan algebras of degree three and their
associated geometries arise naturally within the framework of $N=2$
Maxwell-Einstein supergravity theories (MESGT) in five dimensions
\cite{GST1}. The geometries and symmetries of the corresponding four
dimensional MESGT's are deeply related to the FTS's defined over
these Jordan algebras \footnote{For a review and the references on
the subject see \cite{MG2005}.}. The formally real Jordan algebras
of degree three and the corresponding FTS's were also shown to be
related to relativistic point particle and classical bosonic string
actions by Sierra \cite{Sierra}.

In this paper we will study the quasiconformal groups associated
with FTS's defined over formally real Jordan algebras \footnote{
Formally real or Euclidean Jordan algebras J are such that $a^2 +b^2
=0$ implies that both $a=0$ and $b=0$ for all $ a,b \in J$. } a la
Freudenthal. Such FTS's were classified by Ferrar \cite{ferrar} who
showed that FTS's defined over Jordan algebras require that the
Jordan algebras be of degree three. The generic formally real Jordan
algebras of degree three define  spacetimes which can be interpreted
as extensions of Minkowskian spacetimes by an extra dilatonic
coordinate. Their Lorentz groups are of the form
\begin{equation}
   \mathrm{SO}(d-1,1) \times \mathrm{SO}(1,1)
\end{equation}
and their conformal groups are
\begin{equation}
 \mathrm{SO}(d,2)\times \mathrm{SO}(2,1)
\end{equation}

The spacetimes defined by four simple formally real Jordan  algebras
of degree three describe extensions of the {\it critical}
Minkowskian spacetimes in $d=3,4,6,10$ dimensions by a dilatonic
{\it and} $2,4,8, 16$ spinorial commuting coordinates, respectively.
Rotation groups of these space-times are
\begin{equation}
   \mathrm{SO}\left(3\right), \quad
   \mathrm{SU}\left(3\right), \quad
   \mathrm{USp}\left(6\right), \quad
   \mathrm{F}_4
\end{equation}
Their Lorentz groups are
\begin{equation}
  \mathrm{SL}\left(3,\mathbb{R}\right),  \quad
  \mathrm{SL}\left(3,\mathbb{C}\right), \quad
  \mathrm{SU}^\ast\left(6\right), \quad
  \mathrm{E}_{6(-26)}
\end{equation}
and their conformal groups are
\begin{equation}
 \mathrm{Sp}\left(6, \mathbb{R}\right), \quad
 \mathrm{SU}\left(3,3\right),           \quad
 \mathrm{SO}^\ast\left(12\right), \quad
 \mathrm{E}_{7(-25)}
\end{equation}
respectively. Hence their rotation, Lorentz and conformal groups
coincide with the first three rows of the Magic Square
\cite{FrSprTits}.

The quasiconformal groups associated with the spacetimes defined by
the generic Jordan algebras of degree three are
\begin{equation}
   \mathrm{SO}\left(d+2,4\right)
\end{equation}
while the quasiconformal groups associated with the spacetimes
defined by simple Jordan algebras of degree are
\begin{equation}
  \mathrm{F}_{4(4)}, \quad \mathrm{E}_{6(2)} , \quad
  \mathrm{E}_{7(-5)}, \quad  \mathrm{E}_{8(-24)}
\end{equation}

One of the remarkable features of the quasiconformal realizations of
noncompact groups is that their quantization leads naturally to the
minimal unitary representations of the respective noncompact groups
\cite{GKN2,GP,MG2005}. The concept of a minimal unitary
representation of a non-compact group $G$ was first introduced by
A.~Joseph \cite{Joseph}. Over the last two decades there has been a
great deal of work done by the mathematicians on the minimal unitary
representations of noncompact groups. For references on earlier work
on the subject we refer the reader to the review lectures of
Jian-Shu Li \cite{li2000}. More recently, minimal unitary
representations were studied by Kazhdan, Pioline, and Waldron
\cite{PKW}, by G\"unaydin, Koepsell and Nicolai \cite{GKN2} and by
G\"unaydin and Pavlyk \cite{GP}. The construction of KPW extends to
all simply laced groups and was motivated by the idea that the theta
series of $\mathrm{E}_{8(8)}$ and its subgroups may describe the
quantum supermembrane in various dimensions~\cite{PW}.  One of the
main motivations of the work of GKN as well as of GP was the idea
that the spectra of M-theory in various dimensions must fall into
unitary representations of its U-duality groups in the respective
dimensions. Furthermore the U-duality groups in 3 and 4 dimensions
act as spectrum generating symmetry groups in the charge space of
BPS black hole solutions in 4 and 5 dimensions , respectively
\cite{GKN1,MG2005}.  Realization of the minimal unitary
representation of $\mathrm{E}_{8(8)}$ and its subalgebras given in
\cite{GKN2} is based on their geometric realizations as
quasi-conformal groups.  In our earlier work \cite{GP} we
constructed the minimal unitary representation of the other
noncompact real form of $E_8$, namely $E_{8(-24)}$, and those of its
subgroups, that arise as U-duality groups of MESGT's defined by
simple Jordan algebras of degree three. These minimal unitary
representations correspond to quantizations of their geometric
realizations as quasiconformal groups as well. In this paper we will
study the quasiconformal groups associated with all Euclidean Jordan
algebras of degree three mainly from a space-time point of view. We
will also give the minimal unitary realizations of the
quasiconformal groups $SO(d+2,4)$ of spacetimes defined by the
generic Jordan family.

 The plan of the paper is as follows. We review $U$-duality groups
arising in $N=2$ Maxwell-Einstein supergravity theories (MESGT)
defined by cubic forms in dimensions 5,4 and 3 in section
\ref{sec:Udual}.  In section \ref{sec:G-SpaceTimes} we review
generalized space-times defined by Jordan algebras as well as their
symmetry groups. In section \ref{sec:SpaceTime-Ext} we study the
spacetimes defined by Jordan algebras of degree 3 as extensions of
Minkowskian space-times with dilatonic and spinorial coordinates and
their symmetries. We the review the realization of the
quasiconformal groups via the Freudenthal triple product given in
\cite{GKN1}. In section 5 we present the explicit geometric
realizations of quasi-conformal groups $SO\left(d+2,4\right)$ of the
generic Jordan family of spacetimes. In section 6 we give an
explicit geometric realization of $\mathrm{E}_{8(-24)}$ as a
quasiconformal group as the aforementioned extension of
$\mathrm{SO}\left(12,4\right)$. By consistent truncation of
$E_{8(-24)}$ we obtain the geometric quasiconformal realizations of
$E_{7(-5)}, E_{6(2)}$ and $F_{4(4)}$, which we give explicitly as
well. In the final section we give the minimal unitary realizations
of the generic family of quasiconformal groups
$\mathrm{SO}\left(d+2,4\right)$. Together with our earlier results
given in \cite{GP} this completes the construction of the minimal
unitary representations of quasiconformal groups defined over
Euclidean Jordan algebras of degree three by quantizations of their
geometric realizations.

\section{U-duality groups of $N=2$ Maxwell-Einstein supergravity theories in $d=5,4,3$
dimensions defined by cubic forms \label{sec:Udual}}

\subsection{Five dimensional $N=2$ MESGT's }
In this section we will review the geometry and the symmetry groups of
$N=2$ MESGT's in five dimensions and the corresponding dimensionally
reduced theories in $d=4$ and $d=3$ dimensions.  The MESGT's describe
the coupling of an arbitrary number $n$ of (Abelian) vector fields to
$N=2$ supergravity and five dimensional MESGT's were constructed in
\cite{GST1}. The bosonic part of the Lagrangian can be written as
\cite{GST1}
\begin{eqnarray}
   \label{Lagrange}
e^{-1}\mathcal{L}_{\rm bosonic}&=& -\frac{1}{2}R
-\frac{1}{4}{\stackrel{\circ}{a}}_{IJ}F_{\mu\nu}^{I}
F^{J\mu\nu}-\frac{1}{2}g_{xy}(\partial_{\mu}\varphi^{x})
(\partial^{\mu} \varphi^{y}) \nonumber \\ &&+
 \frac{e^{-1}}{6\sqrt{6}}C_{IJK}\varepsilon^{\mu\nu\rho\sigma\lambda}
 F_{\mu\nu}^{I}F_{\rho\sigma}^{J}A_{\lambda}^{K},
\end{eqnarray}
where $e$ and $R$ denote the f\"{u}nfbein determinant and the scalar
curvature in $d=5$, respectively. $F_{\mu\nu}^{I}$ are the field
strengths of the Abelian vector fields $A_{\mu}^{I}, \,( I=0,1,2
\cdots, n$) with $A^0_{\mu}$ denoting the ``bare'' graviphoton. The
metric, $g_{xy}$, of the scalar manifold $\mathcal{M}$ and the
``metric'' ${\stackrel{\circ}{a}}_{IJ}$ of the kinetic energy term of
the vector fields both depend on the scalar fields $\varphi^{x}$ (
$x,y,..=1,2,..,n$). The invariance under Abelian gauge transformations
of the vector fields requires the completely symmetric tensor
$C_{IJK}$ to be constant. Remarkably, one finds that the entire $N=2$,
$d=5$ MESGT is uniquely determined by the constant tensor $C_{IJK}$
\cite{GST1}. In particular, the metrics of the kinetic energy terms
of the vector and scalar fields are determined by $C_{IJK}$. More
specifically, consider the cubic polynomial, $\mathcal{V}(h)$, in
$(n+1)$ real variables $h^{I}$ $(I=0,1,\ldots,n)$ defined by the
$C_{IJK}$
\begin{equation}
\mathcal{V}(h):=C_{IJK} \, h^{I} h^{J} h^{K}\ .
\end{equation}
Using this polynomial as a real ``K\"ahler potential'' for a metric,
$a_{IJ}$, in an $ n+1 $ dimensional ambient space with the coordinates
$h^{I}$:
\begin{equation}\label{aij}
a_{IJ}(h):=-\frac{1}{3}\frac{\partial}{\partial h^{I}}
\frac{\partial}{\partial h^{J}} \ln \mathcal{V}(h)\ .
\end{equation}
one finds that the $n$-dimensional target space, $\mathcal{M}$, of the
scalar fields $\varphi^{x}$ can be identified with the
hypersurface~\cite{GST1}
\begin{equation}
  \mathcal{V} (h)=C_{IJK}h^{I}h^{J}h^{K}=1
\end{equation}
in this  space. The metric  $g_{xy}$ of the scalar manifold is
simply  the pull-back of (\ref{aij}) to $\mathcal{M}$
\begin{equation}
  g_{xy} = h^I_x h^J_y {\stackrel{\circ}{a}}_{IJ}
\end{equation}
where
\begin{equation}
 h^I_x = -\sqrt{\frac{3}{2}} \frac{\partial}{\partial\phi^x} h^I
\end{equation}
and one finds that the Riemann curvature of the scalar manifold
has the simple form
\begin{equation}
  K_{xyzu}= \frac{4}{3} \left( g_{x[u} g_{z]y} + {T_{x[u}}^w T_{z]yw} \right)
\end{equation}
where $T_{xyz}$ is  the symmetric tensor
\begin{equation}
   T_{xyz}= h^I_x h^J_y h^K_z C_{IJK}
\end{equation}
The ``metric'' ${\stackrel{\circ}{a}}_{IJ}(\varphi)$
of the kinetic energy term of the vector fields appearing in
\eqref{Lagrange} is given by the componentwise restriction of
$a_{IJ}$ to $\mathcal{M}$:
\begin{equation}
{\stackrel{\circ}{a}}_{IJ}\left(\varphi\right) = \left. a_{IJ}\right|_{\mathcal{V}=1} \ .
\end{equation}
We should stress that the indices $I,J,K,..$ are  lowered and raised
by the metric $ {\stackrel{\circ}{a}}_{IJ}(\varphi) $ and its
inverse. The physical requirement of positivity of kinetic energy
requires that $g_{xy}$ and ${\stackrel{\circ}{a}}_{IJ}$ be positive
definite metrics. This requirement induces constraints on the
possible $C_{IJK}$, and in \cite{GST1} it   was shown that any
$C_{IJK}$ that satisfy these constraints can be brought to the
following form
\begin{equation}
   \label{canbasis}
   C_{000}=1,\quad C_{0ij}=-\frac{1}{2}\delta_{ij},\quad  C_{00i}=0,
\end{equation}
with the remaining coefficients $C_{ijk}$ ($i,j,k=1,2,\ldots, n$)
being completely arbitrary.  This basis is referred to as the
canonical basis for $C_{IJK}$.

Denoting the symmetry group of the tensor $C_{IJK}$ as $G$ one finds
that the full symmetry group of $N=2$ MESGT in $d=5$ is of the form $
G \times \mathrm{SU}(2)_R $ where $\mathrm{SU}(2)_R$ denotes the local
R-symmetry group of the $N=2$ supersymmetry algebra.

\subsection{Symmetric target spaces and Jordan Algebras\label{sec:review}}

From the form of the Riemann curvature tensor $K_{xyzu}$ it is clear
that the covariant constancy of $T_{xyz}$ implies the covariant
constancy of $K_{xyzu}$:
\begin{equation*}
   T_{xyz; w} = 0  \implies  K_{xyzu ; w} =0
\end{equation*}
Therefore the scalar manifolds $\mathcal{M}_5$ with covariant
 constantly constant $T$ tensor are locally symmetric spaces.

If $\mathcal{M}_5$ is a homogeneous space the covariant constancy of
$T_{xyz}$ was shown to be equivalent to the following
identity~\cite{GST1}:
\begin{equation}
   C^{IJK} C_{J(MN} C_{PQ)K} = {\delta^I}_{(M} C_{NPQ)}
\end{equation}
where the indices are raised by ${\stackrel{\circ}{a}}{}^{IJ}$.\footnote{ 
For proof of this equivalence an expression for constants $C_{IJK}$
in terms of scalar field dependent quantities was used
\begin{equation*}
   C_{IJK} = \frac{5}{2} h_I h_K h_K - \frac{3}{2} {\stackrel{\circ}{a}}_{(IJ} h_{K)} +
    T_{xyz} h^x_I h^y_J h^z_K
\end{equation*}
as well as algebraic constraints $h_I h^I=1$ and $h^I_x h_I =0$ that
follows from susy.  }

Remarkably the cubic forms defined by $C_{IJK}$ of the $N=2$ MESGT's
with $n \geq 2$ with a symmetric target space $\mathcal{M}_5$ and a
covariantly constant $T$ tensor are in one-to-one correspondence
with the norm forms of Euclidean (formally real) Jordan algebras of
degree 3.

The precise connection between Jordan algebras of degree 3 and the
geometries of MESGT's with symmetric target spaces in $d=5$ was
established \cite{GST1} through a novel formulation of the
corresponding Jordan algebras. This formulation is due to McCrimmon
\cite{McCrimmon}, who generalized and unified previous constructions
by Freudenthal, Springer and Tits \cite{FrSprTits}, which we outline
here.

Let $V$ be a vector space over the field of reals $\mathbb{R}$, and
let $\mathcal{V} \colon V \times V \times V \to \mathbb{R}$ be a cubic
norm on $V$. Furthermore, assume that there exists a quadratic map
$\sharp: x \to x^\sharp$ of $V$ into itself and a ``base point'' $c
\in V$ such that
\begin{subequations}
\begin{equation}
  \mathcal{V}\left(c\right) = 1 \phantom{and also}  \text{and} \phantom{and also} c^\sharp = c   \tag*{(i), (ii)}
\end{equation}
\begin{equation}
  T\left(x^\sharp, y\right)  = \left. y^I \partial_I \mathcal{V}
  \right|_x
   \tag{iii}
\end{equation}
\begin{equation}
   c \times y = T\left(y, c\right) c - y   \tag{iv}
\end{equation}
\begin{equation}
   \left(x^\sharp\right)^\sharp = \mathcal{V}\left(x\right) x
   \tag{v}
\end{equation}
 The last equation is  referred to as the adjoint identity. The map  $T: V \times V \to \mathbb{R}$
 is defined as
\begin{equation}
   T\left(x, y\right) = - \left. x^I y^J \partial_I \partial_J \ln \mathcal{V} \right|_c
\end{equation}
and
 the Freudenthal product $\times $ of two elements $x$ and $y$ is defined as
\begin{equation}
  x \times y = \left(x+y\right)^\sharp - x^\sharp - y^\sharp
\end{equation}
\end{subequations}

McCrimmon showed that a vector space with the above properties
defines a unital Jordan algebra with Jordan product $\circ$ given by
\begin{equation}
   x \circ y = \frac{1}{2} \Big( T\left(c,x\right) y + T\left(c, y\right) x -
           T \left(c, x \times y\right) c + x \times y \Big)
\end{equation}
and a quadratic operator $U_x$ given by
\begin{equation}\label{eq:Uxtimesy}
  U_x y = T\left(x, y\right) x - x^\sharp \times y
\end{equation}
In \cite{GST1} it was shown that the properties (i) and (iv) are
satisfied by the cubic norm form defined by the tensor $C_{IJK}$ of
$N=2$  MESGT's in $d=5$. The condition of adjoint identity is
equivalent to the requirement that the scalar manifold be symmetric
space with a covariantly constant $T$-tensor \cite{GST1}. The
corresponding symmetric spaces are of the form
\begin{equation}
    \mathcal{M} = \frac{\mathrm{Str}_0 \left(J\right)}{ \mathrm{Aut}\left(J\right)}
\end{equation}
where $\mathrm{Str}_0\left(J\right)$ and
$\mathrm{Aut}\left(J\right)$ are the reduced structure group and
automorphism group of the Jordan algebra $J$ respectively.

 From the foregoing we see that the classification of locally
symmetric spaces $\mathcal{M}$ for which the tensor $T_{xyz}$ is
covariantly constant reduces to the classification of Jordan
algebras with cubic norm forms. Following  Schafers \cite{Schafers}
the possibilities were listed in \cite{GST1}:
\begin{enumerate}
   \item $J= \mathbb{R}$, $\mathcal{V}\left(x\right) = x^3$. The base point may be chosen as $c=1$. This
case supplies $n=0$, i.e. pure $d=5$ supergravity.
   \item $J = \mathbb{R} \oplus \Gamma$, where $\Gamma$ is a simple algebra with identity $\mathbf{e}_2$ and
quadratic norm $Q\left(\mathbf{x}\right)$, for $\mathbf{x} \in
\Gamma$, such that $Q\left(\mathbf{e}_2\right)=1$.
 The norm is given as $\mathcal{V}\left(x\right) = a Q\left(\mathbf{x}\right)$, with $x=\left(a, \mathbf{x}\right)$.
 The base point may be chosen as $c = \left(1, \mathbf{e}_2\right)$. This includes two special cases
\begin{enumerate}
  \item $\Gamma = \mathbb{R}$ and $Q = b^2$, with $\mathcal{V} = a b^2$. This is applicable to $n=1$.
  \item $\Gamma = \mathbb{R} \oplus  \mathbb{R}$ and $Q=bc$, and $\mathcal{V} = a b c$ and is applicable to $n=2$.
\end{enumerate}
 Notice that for these special cases the norm is completely factorized, so that the space $\mathcal{C}$ and therefore
$\mathcal{M}$, is flat. For $n>2$, $\mathcal{V}$ is still factorized
into a linear and quadratic parts, so that $ \mathcal{M}$ is still
reducible. The positive definiteness of the metric $a_{IJ}$ of
$\mathcal{C}$, which is required on the physical grounds, requires
that $Q$ have Minkowski signature $\left(+,-,-,\ldots,-\right)$. The
point $\mathbf{e}_2$ can be chosen as $\left(1, 0, \ldots,
0\right)$. It is then obvious that the invariance group of the norm
is
 \begin{equation}
    \mathrm{Str}_0\left(J\right) = \mathrm{SO}\left(n-1, 1\right) \times \mathrm{SO}\left(1, 1\right)
 \end{equation}
where the $\mathrm{SO}\left(1, 1\right)$ factor arises from the
invariance of $\mathcal{V}$ under the dilatation $\left(a,
\mathbf{x}\right)
 \to \left( e^{-2\lambda} a, e^{\lambda} \mathbf{x}\right)$ for $\lambda \in \mathbb{R}$, and that
$\mathrm{SO}\left(n-1\right)$ is $\mathrm{Aut}\left(J\right)$. Hence
\begin{equation}
    \mathcal{M} = \frac{\mathrm{SO}\left(n-1,1\right)}{\mathrm{SO}\left(n-1\right)} \times \mathrm{SO}\left(1,1\right)
\end{equation}

\item Simple Euclidean Jordan algebras  $J = J_3^\mathbb{A}$ generated by $3\times 3$ Hermitian matrices
over the four division algebras  $\mathbb{A}= \mathbb{R}$,
$\mathbb{C}$, $\mathbb{H}$, $\mathbb{O}$. In these four cases an
element $x \in J$ can be written as
\begin{equation}
   x = \begin{pmatrix}
         \alpha_1 &  a_3   & a^\ast_2 \cr
         a^\ast_3 & \alpha_2 & a_1 \cr
          a_2 & a^\ast_1 & \alpha_3
     \end{pmatrix}
\end{equation}
where $\alpha_k \in \mathbb{R}$ and $a_k \in \mathbb{A}$ with $*$
indicating the conjugation in the underlying dvision algebra. The
cubic norm $\mathcal{V}$, following Freudenthal \cite{FrSprTits},
is given by
\begin{equation}
   \mathcal{V}\left(x\right) = \alpha_1\alpha_2\alpha_3 - \alpha_1 \left|{a_1}\right|^2 - \alpha_2 \left|{a_2}\right|^2
     - \alpha_3 \left|{a_3}\right|^2 + a_1 a_2 a_3 + \left( a_1 a_2 a_3 \right)^\ast
\end{equation}
For $\mathbb{A}=\mathbb{R}$ or $\mathbb{C}$ it coincides with the
usual definition of determinant $Det (x)$. The corresponding spaces
$\mathcal{M}$ are irreducible of dimension $3 \left(1+\dim
\mathbb{A}\right)-1$, which we list below:
\begin{equation}
    \begin{array}{cc}
        \mathcal{M}( J_3^\mathbb{R}) =\phantom{gis }  & \dfrac{\mathrm{SL}\left(3, \mathbb{R}\right)}
        {\mathrm{SO}\left(3\right)} \\[12pt]
         \mathcal{M}(J_3^\mathbb{C} ) = \phantom{ges } & \dfrac{\mathrm{SL}\left(3, \mathbb{C}\right)}{\mathrm{SU}\left(3\right)}
    \end{array}
    \phantom{ and also }
    \begin{array}{cc}
         \mathcal{M}(J_3^\mathbb{H}) =\phantom{ges }  & \dfrac{\mathrm{SU}^\ast\left(6\right)}{\mathrm{USp}\left(6\right)} \\[12pt]
         \mathcal{M}(J_3^\mathbb{O}) =\phantom{ges } & \dfrac{\mathrm{E}_{6(-26)}}{\mathrm{F}_4}
    \end{array}
\end{equation}

\end{enumerate}


 The magical supergravity theories described by simple Jordan
algebras $J_A^\mathbb{A}$ ($\mathrm{A}$ = $\mathrm{R}$,
$\mathrm{C}$, $\mathrm{H}$ or $\mathrm{O}$) can be truncated to
theories belonging to the generic families. This is achieved by
restricting the elements of $J_A^\mathbb{A}$
\begin{equation}
    \begin{pmatrix}
         \alpha_1 &  a_3   & \overline{a}_2 \cr
         \overline{a}_3 & \alpha_2 & a_1 \cr
          a_2 & \overline{a}_1 & \alpha_3
    \end{pmatrix}
\end{equation}
to lie in their subalgebra $J = \mathbb{R} \oplus J_2^\mathbb{A}$ be setting $a_1=a_2=0$. Their symmetry groups
are as follows:
\begin{equation}
\begin{split}
    J = \mathbb{R} \oplus J_2^\mathbb{R}  &: \mathrm{SO}(1,1) \times \mathrm{SO}\left(2,1\right) \subset \mathrm{SL}\left(3, \mathbb{R}\right) \cr
    J = \mathbb{R} \oplus J_2^\mathbb{C}  &: \mathrm{SO}(1,1) \times \mathrm{SO}\left(3,1\right) \subset \mathrm{SL}\left(3, \mathbb{C}\right) \cr
     J = \mathbb{R} \oplus J_2^\mathbb{H}  &: \mathrm{SO}(1,1) \times \mathrm{SO}\left(5,1\right) \subset \mathrm{SU}^\ast\left(6\right) \cr
     J = \mathbb{R} \oplus J_2^\mathbb{O}  &: \mathrm{SO}(1,1) \times \mathrm{SO}\left(9,1\right) \subset \mathrm{E}_{6(-26)}
\end{split}
\end{equation}

\subsection{Geometries of the four dimensional MESGTs defined by
Jordan algebras of degree 3}

Under dimensional reduction to the $4D$ the kinetic energy of the scalar fields of the $5D$ $N=2$ MESGTs can be
written as \cite{GST1}
\begin{equation}
   e^{-1} \mathcal{L}_\mathrm{scalars} = - g_{IJ} \partial_\mu Z^I \partial^\mu \overline{Z}^J
\end{equation}
where
\begin{equation}
    g_{IJ} = \Hat{a}_{IJ} \left(Z - \overline{Z}\right) = - \frac{1}{2} \frac{\partial}{\partial Z^I}
\frac{\partial}{\partial \overline{Z} {}^J} \ln \mathcal{V}\left(Z - \overline{Z}\right)
\end{equation}
and $Z^I$ are complex scalar fields
\begin{equation}
   Z^I = \frac{1}{\sqrt{2}} \left( \sqrt{ \frac{2}{3} } A^I + i \Hat{h}^I \right)
\end{equation}
where the real parts $A^I$  are scalars coming from the vectors in 5
dimensions and $\Hat{h}^I$ are
\begin{equation}
   \Hat{h}^I = e^{\sigma} h^I\left( \phi^x\right)
\end{equation}
where $\sigma$ is the scalar coming from the graviton in the five dimensions.
Since $\mathcal{V}\left(\Hat{h}\right) = e^{3\sigma} > 0$ the scalar
manifold in $4D$ theories corresponds to the ``upper half-plane''
with respect to the cubic norm. For Euclidean Jordan algebras of
degree three these are the Koecher upper half-spaces \cite{Koecher}
of the corresponding Jordan algebras
\begin{equation}
    \mathcal{M}_4 = \mathcal{D} \left(J\right) = J + i \mathcal{C}\left(J\right)
\end{equation}
where $\mathcal{C}\left(J\right)$ denotes elements of the Jordan
algebra with positive cubic norm. The Koecher half-spaces are
bi-holomorphically equivalent to bounded symmetric domains whose
Bergmann kernel is simply $\mathcal{V}\left(Z-\overline{Z}\right)$.
As was first shown in \cite{four} the scalar manifold of the $4D$
MESGTs must be special K\"ahler. For the theories coming from $5D$
the K\"ahler potential reads
\begin{equation}
  F\left(Z, \overline{Z}\right) = -\frac{1}{2} \ln \mathcal{V} \left( Z - \overline{Z}\right)
\end{equation}
and are called very special K\"ahler geometries.

The bounded symmetric domains associated with the upper half-spaces of Jordan algebras are
isomorphic to certain hermitian symmetric spaces. For the Euclidean Jordan algebras of degree 3 these
spaces are as follows:
\begin{equation}
\begin{split}
    \mathcal{M}_4 \left( J = \mathbb{R} + \Gamma\left(Q\right)\right) &= \frac{
     \mathrm{SO}\left(2,1\right) \times \mathrm{SO}\left(n,2\right)
   }{  \mathrm{SO}(2) \times \mathrm{SO}\left(n\right) \times \mathrm{SO}(2)}  \cr
\mathcal{M}_4 \left( J_3^\mathbb{R} \right) &= \frac{
\mathrm{Sp}\left(6, \mathbb{R}\right)}{\mathrm{U}(3)} \cr
   \mathcal{M}_4 \left( J_3^\mathbb{C} \right) &=
        \frac{ \mathrm{SU}\left(3, 3\right)}{ \mathrm{S} \left( \mathrm{U}(3) \times \mathrm{U}(3) \right)} \cr
   \mathcal{M}_4 \left( J_3^\mathbb{H} \right) &= \frac{ \mathrm{SO}^\ast\left(12\right)}{\mathrm{U}(6)} \cr
   \mathcal{M}_4 \left( J_3^\mathbb{O} \right) &= \frac{ \mathrm{E}_{7(-25)}}{ \mathrm{E}_6 \times \mathrm{U}(1)} \cr
\end{split}
\end{equation}
These symmetric spaces are simply the quotients of the conformal
groups of the corresponding Jordan algebras  by their maximal
compact subgroups:
\begin{equation*}
    \mathcal{M}_4 = \frac{\mathrm{Conf}\left(J\right)}{ K \left(J\right) }
\end{equation*}
The correspondence between the vector fields and the elements of the
underlying Jordan algebras in five dimensions gets extended to a
correspondence between the vector field strengths $F_{\mu\nu}^{A}$
plus  their magnetic duals $G_{\mu\nu}^A$ with the elements of the
Freudenthal triple system defined by the Jordan algebra of degree
three
\begin{equation}
 F_{\mu\nu}^{A} \oplus G_{\mu\nu}^A \Leftrightarrow \mathcal{FTS} (J)
\end{equation}

The automorphism group of this FTS is isomorphic to the four
dimensional U-duality group and it acts as the spectrum generating
conformal group on the charge space of the BPS black hole solutions
of  five dimensional MESGT's \cite{GKN1,MG2005}.

\subsection{Geometries of the three  dimensional MESGTs defined by Jordan algebras of degree 3}

Upon further dimensional reduction to 3 space-time dimensions, the
MES\-GTs defined by Euclidean Jordan algebras of degree three have
target spaces that are quaternionic symmetric spaces. The
corresponding symmetric spa\-ces are:
\begin{equation}
\begin{split}
    &\phantom{this is centerd}\mathcal{M}_3 \left( J = \mathbb{R} + \Gamma\left(Q\right)\right) = \frac{
     \mathrm{SO}(n+2,4)}{\mathrm{SO}(n+2) \times \mathrm{SO}(4) } \cr
  &\begin{aligned}
  \mathcal{M}_3\left( J_3^\mathbb{R} \right) &= \frac{
  \mathrm{F}_{4(4)}}{\mathrm{USp}(6) \times \mathrm{SU}(2)} \cr
   \mathcal{M}_3 \left( J_3^\mathbb{C} \right) &=
        \frac{ \mathrm{E}_{6(2)}}{ \mathrm{SU}(6)\times \mathrm{SU}(2)}
  \end{aligned}
   \qquad
   \begin{aligned}
   \mathcal{M}_3 \left( J_3^\mathbb{H} \right) &= \frac{ \mathrm{E}_{7(-5)}}{\mathrm{SO}(12) \times SU(2)} \cr
   \mathcal{M}_3 \left( J_3^\mathbb{O} \right) &= \frac{ \mathrm{E}_{8(-24)}}{ \mathrm{E}_7 \times \mathrm{SU}(2)}
   \end{aligned}
\end{split}
\end{equation}
The pure $5d$, $N=2$ supergravity under dimensional reduction to three dimensions leads to the target space
\begin{equation}
  \frac{\mathrm{G}_{2(2)}}{\mathrm{SU}(2)\times \mathrm{SU}(2)}
\end{equation}
which can be embedded in the coset space
\begin{equation}
  \frac{\mathrm{SO}(3,4)} {\mathrm{SO}(3) \times \mathrm{SO}(4)}
\end{equation}

We should note that the above target spaces are obtained after
dualizing all the bosonic propagating fields to scalar fields which
is special to three dimensions. The Lie algebras of the three
dimensional U-duality groups have a 5-graded decomposition with
respect to the four dimensional  U-duality groups. They are
isomorphic to the quasiconformal groups constructed over the
corresponding FTS's, which act as spectrum generating symmetry group
on the charge-entropy space of BPS black hole solutions in four
dimensional MESGT's \cite{GKN1,MG2005}.

\section{Generalized space-times  defined by Jordan  algebras and their
  symmetry groups  \label{sec:G-SpaceTimes}}
\subsection{Generalized Rotation, Lorentz and Conformal\\ Groups}

In the previous sections we reviewed how Jordan algebras arise in a
fundamental way within the framework of supergravity theories. In
this section we will review work on how Jordan algebras can be used
to generalize four dimensional Minkowski spacetime and its symmetry
groups in a natural way. The first proposal to use Jordan algebras
to define generalized spacetimes was made in the early days of
spacetime supersymmetry in attempts to find the super analogs of the
exceptional Lie algebras \cite{mg75}.  One of the main anchors of
this proposal was the  twistor formalism, which in four-dimensional
space-time $(d=4)$ leads naturally to the representation of four
vectors in terms of $2\times 2$ Hermitian matrices over the field of
complex numbers ${\mathbb C}$. In particular, a coordinate
four-vector $x_{\mu}$ can be represented as:
\begin{equation}
    x=x_{\mu}\sigma^{\mu}
\end{equation}
Since the Hermitian matrices over the field of complex numbers close
under the symmetric anti-commutator product one can regard the
coordinate vectors as elements of a Jordan algebra denoted as
$J_2^{\mathbb C}$ \cite{mg75,mg80}.  Then the rotation, Lorentz and
conformal groups in $d=4$ can be identified with the automorphism,
reduced structure and M\"{o}bius (linear fractional) groups of the
Jordan algebra of $2\times 2$ complex Hermitian matrices $J_2^{\mathbb
C}$ \cite{mg75,mg80}.  The reduced structure group $Str_0(J)$ of a
Jordan algebra $J$ is simply the invariance group of its norm form
$N(J)$. (The structure group $Str(J)= Str_0(J)\times SO(1,1)$, on the
other hand, is simply the invariance group of $N(J)$ up to an overall
constant scale factor.) Furthermore, this interpretation leads one
naturally to define generalized space-times whose coordinates are
parameterized by the elements of Jordan algebras \cite{mg75}. The
rotation $Rot(J)$, Lorentz $Lor(J)$ and conformal $Con(J)$ groups of
these generalized space-times are then identified with the
automorphism $Aut(J)$, reduced structure $Str_0(J)$ and M\"{o}bius
M\"{o}(J) groups of the corresponding Jordan algebras
\cite{mg75,mg80,mg91,mg92}.  Denoting as $J_{n}^{\mathbb A}$ the
Jordan algebra of $n\times n$ Hermitian matrices over the {\it
division} algebra ${\mathbb A}$ and the Jordan algebra of Dirac gamma
matrices in $d$ (Euclidean) dimensions as $\Gamma(d)$ one finds the
following symmetry groups of generalized space-times defined by simple
Euclidean (formally real) Jordan algebras:
\begin{center}
\begin{tabular}{|c|c|c|c|}
 \hline
$J $& $Rotation(J)$ & $Lorentz(J)$ & $Conformal(J)$ \\
\hline
~&~&~&~\\
$J_{n}^{\mathbb R}$ & $SO(n)$ & $SL(n,{\mathbb R})$ & $Sp(2n,{\mathbb R})$\\
~&~&~&~\\
$J_{n}^{\mathbb C}$ & $SU(n) $ & $ SL(n,{\mathbb C})$ & $SU(n,n)$ \\
~&~&~&~\\
$J_{n}^{\mathbb H} $& $USp(2n)$ & $SU^{*}(2n)$ &$ SO^{*}(4n)$ \\
~&~&~&~\\
$J_{3}^{\mathbb O}$ &$ F_{4}$ & $E_{6(-26)}$ & $E_{7(-25)}$ \\
~&~&~&~ \\
$\Gamma(d)$ & $SO(d)$ & $SO(d,1)$ & $SO(d,2)$ \\
~&~&~&~ \\
\hline
\end{tabular}

\end{center}

The symbols ${\mathbb R}$, ${\mathbb C}$, ${\mathbb H}$, ${\mathbb O}$
represent the four division algebras.  For the Jordan algebras
$J_n^{\mathbb A}$ the norm form is the determinental form (or its
generalization to the quaternionic and octonionic matrices). For the
Jordan algebra $\Gamma(d)$ generated by Dirac gamma matrices
$\Gamma_{i} ~( i =1,2,...d)$

\begin{equation}
~\\~ \{\Gamma_i,\Gamma_j\} = \delta_{ij} {\mathbf 1};
~~~~~i,j,\ldots~=~ 1,2,\ldots,d \\
~\\
\end{equation}
the norm of a general element $x= x_0 {\mathbf 1} + x_i \Gamma_i$ of
$\Gamma(d)$ is quadratic and given by
\begin{equation}
  N(x) = x \bar{x}= x_0^2 - x_i x_i
\end{equation}
 where $\bar{x}= x_0 {\mathbf 1} - x_i \Gamma_i $. Its automorphism,
reduced structure and M\"{o}bius groups are simply the rotation,
Lorentz and conformal groups of $(d+1)$-dimensional Minkowski
spacetime. One finds the following special isomorphisms between the
Jordan algebras of $2\times 2$ Hermitian matrices over the four
division algebras and the Jordan algebras of gamma matrices:

\begin{equation}
~\\
J_{2}^{\mathbb R} \simeq \Gamma(2)~~~~;~~~~J_{2}^{\mathbb C} \simeq \Gamma(3) \\
~~~~;~~~~
J_{2}^{\mathbb H} \simeq \Gamma(5)~~~~;~~~~J_{2}^{\mathbb O} \simeq \Gamma(9) \\
~\\
\end{equation}

The Minkowski spacetimes they correspond to are precisely the critical
dimensions for the existence of super Yang-Mills theories as well as
of the classical Green-Schwarz superstrings. These Jordan algebras are
all quadratic and their norm forms are precisely the quadratic
invariants constructed using the Minkowski metric.

We should note two remarkable facts about the above table. First, the
conformal groups of generalized space-times defined by Euclidean
(formally real) Jordan algebras all admit positive energy unitary
representations\footnote{Similarly, the generalized conformal groups
defined by Hermitian Jordan triple systems all admit positive energy
unitary representations \cite{mg92}. In fact the conformal groups of
simple Hermitian Jordan triple systems exhaust the list of simple
noncompact groups that admit positive energy unitary
representations. They include the conformal groups of simple Euclidean
Jordan algebra since the latter form an hermitian Jordan triple system
under the Jordan triple product \cite{mg92}. }. Hence they can be
given a causal structure with a unitary time evolution as in four
dimensional Minkowski space-time. Second is the fact that the maximal
compact subgroups of the generalized conformal groups of formally real
Jordan algebras are simply the compact forms of their structure groups
(which are the products of their generalized Lorentz groups with
dilatations).

\subsection{Quasiconformal groups and Freudenthal triple systems}

A Freudenthal triple system (FTS) is a vector space $\mathfrak{M}$
with a trilinear product $(X,Y,Z)$ and a skew symmetric bilinear
form $<X,Y>$ such that\footnote{We should note that the triple
product \eqref{ftp56-rel} could be modified by terms involving the
symplectic invariant, such as $\langle X,Y \rangle Z$.  The  choice
given above was made in \cite{GKN1} in order to obtain agreement
with the formulas of~\cite{Faul71}.}:
\begin{eqnarray}
\FTP{X}{Y}{Z} &=& \FTP{Y}{X}{Z} +2\,\Sympl{X}{Y}Z \,,\nonumber \\[1ex]
\FTP{X}{Y}{Z} &=& \FTP{Z}{Y}{X} -2\,\Sympl{X}{Z}Y \,,\nonumber \\[1ex]
\Sympl{\FTP{X}{Y}{Z}}{W} &=& \Sympl{\FTP{X}{W}{Z}}{Y}
                                -2\,\Sympl{X}{Z}\Sympl{Y}{W} \,,\nonumber \\[1ex]
\FTP{X}{Y}{\FTP{V}{W}{Z}} &=& \FTP{V}{W}{\FTP{X}{Y}{Z}}
                             +\FTP{\FTP{X}{Y}{V}}{W}{Z} \nonumber \\
                          && {}+\FTP{V}{\FTP{Y}{X}{W}}{Z} \,.\label{ftp56-rel}
\end{eqnarray}
A  quartic invariant $\mathcal{I}_4$ can be constructed over the FTS by
means of the  triple product and the bilinear form  as
\begin{eqnarray}
\mathcal{I}_4(X) := \Sympl{\FTP{X}{X}{X}}{X} \label{e7-invariant}
\end{eqnarray}
One can construct a Lie algebra over a FTS that has a 5-graded
decomposition such that grade $\pm 2$ subspaces are one dimensional:
\begin{eqnarray}
\mathfrak{g} = \mathfrak{g}^{-2} \oplus \mathfrak{g}^{-1} \oplus
       \mathfrak{g}^0 \oplus \mathfrak{g}^{+1}
      \oplus \mathfrak{g}^{+2} \,.
\end{eqnarray}
Following \cite{GKN1} we shall label the  Lie algebra generators
belonging to grade $+1$ and grade $-1$ subspaces as $U_A$ and
$\tilde U_A$, where $A \in \mathfrak{M}$. The generators $S_{AB}$
belonging to grade zero subspace are  labeled by a pair of elements
$A,B \in \mathfrak{M}$. For the grade $\pm 2$ subspaces one would in
general need another set of generators $K_{AB}$ and $\tilde K_{AB}$
labeled by two elements, but since these subspaces are
one-dimensional we can write them as
\begin{equation}
  K_{AB} = \left<A,B\right> K_a \qquad\quad
  \tilde{K}_{AB} =\left<A,B\right> \tilde{K}_a
\end{equation}
where a is a real parameter.

One can realize the Lie algebra $\mathfrak{g}$ as a quasiconformal Lie
algebra over a vector space whose coordinates $\cX$ are labeled by a
pair $\left(X,x\right)$, where $X \in \mathfrak{M}$ and $x$ is an
extra single variable as follows \cite{GKN1}:
\begin{equation}
\begin{split}
  \begin{aligned}
      K_a\left(X\right) &= 0 \\
      K_a\left(x\right) &= 2\, a
  \end{aligned}
  & \quad
  \begin{aligned}
     U_A \left(X\right) &= A \\
     U_A\left(x\right) &= \left< A, X\right>
  \end{aligned}
   \quad
   \begin{aligned}
      S_{AB}\left(X\right) &= \left( A, B, X\right) \\
      S_{AB}\left(x\right) &= 2 \left< A, B\right> x
   \end{aligned}
 \\
 &\begin{aligned}
    \Tilde{U}_A\left(X\right) &= \frac{1}{2} \left(X, A, X\right) - A x \\
    \Tilde{U}_A\left(x\right) &= -\frac{1}{6} \left< \left(X, X, X\right), A \right> + \left< X, A\right> x
 \end{aligned}
 \\
 &\begin{aligned}
    \Tilde{K}_a\left(X\right) &= -\frac{1}{6} \,a  \left(X,X,X\right) + a X x \\
    \Tilde{K}_a\left(x\right) &= \frac{1}{6} \, a \left< \left(X, X, X\right), X \right> + 2\, a \, x^2
 \end{aligned}
\end{split}
\end{equation}

From these formulas it is straightforward to determine the
commutation relations of the transformations. The quasiconformal
groups leave invariant a suitable defined lightcone with respect to
a quartic norm involving the quartic invariant of $\mathfrak{M}$
\cite{GKN1}.

Freudenthal introduced the triple systems associated with his name in
his study of the meta\-symplectic geometries associated with exceptional
groups \cite{Freu85}. The geometries associated with FTSs were
further studied in \cite{AllFau84,Faul71,Meyb68,KanSko82}.  A
classification of FTS's may be found in~\cite{KanSko82}, where it is
also shown that there is a one-to-one correspondence between simple
Lie algebras and simple FTS's with a non-degenerate skew symmetric
bilinear form.  Hence there is a quasiconformal realization of every
Lie group acting on a generalized lightcone.

The Freudenthal triple systems associated with exceptional groups
can be represented by formal   $2\times 2$ ``matrices'' of
the form
\begin{equation}
A = \left(
\begin{array}{cc}
  \alpha_1 & x_1 \\ x_2 & \alpha_2
\end{array}   \right)\,,
\end{equation}
where $\alpha_1,\alpha_2$ are real numbers and $x_1,x_2$ are elements
of a simple Jordan algebra $J$ of degree three. One can define a
triple product over the space of such formal matrices such that they
close under it.  There are only four simple Euclidean Jordan algebras
$J$ of this type, namely the $3 \times 3$ hermitean matrices over the
four division algebras, $\mathbb{R}$, $\mathbb{C}$, $\mathbb{H}$ and
$\mathbb{O}$. We shall denote the corresponding FTS's as
$\mathfrak{M}(J)$.

One may ask which Freudenthal triple systems can be realized in the
above form in terms of an underlying Jordan algebra. This question
was investigated by Ferrar \cite{ferrar} who proved that such a
realization is possible only if the underlying Jordan algebra is of
degree three. Remarkably, if one further requires that the
underlying Jordan algebra be formally real then the list of Jordan
algebras over which FTS's can be defined as above coincides with the
list of Jordan algebras that occur in five dimensional $N=2$ MESGT's
whose target spaces are symmetric spaces of the form $G/H$ such that
$G$ is a symmetry of the Lagrangian.

In this paper we will focus only on the quasiconformal groups
defined over formally real Jordan algebras. The Freudenthal triple
product of the elements of $\mathfrak{M}(J)$ is defined as
\cite{Faul71}
\begin{equation}
   \left< X_1, X_2, X_3 \right> = \begin{pmatrix}
      \gamma & c \cr
      d  & \delta
 \end{pmatrix}
\quad \text{with} \quad  X_i = \begin{pmatrix} \alpha_i & a_i \cr b_i & \beta_i \end{pmatrix}
\end{equation}
where
\begin{equation*}
\begin{split}
    \gamma &= \alpha_1 \beta_2 \alpha_3 + 2 \alpha_1 \alpha_2 \beta_3 - \alpha_3 T\left(a_1, b_2 \right)
     - \alpha_2 T\left(a_1, b_3\right) \\
        & - \alpha_1 T\left(a_2, b_3 \right) + T\left( a_1, a_2 \times a_3 \right) \\
    c &= \left( \alpha_2 \beta_3 + T\left(b_2, a_3 \right) \right) a_1 +
        \left( \alpha_1 \beta_3 + T\left(b_1, a_3 \right) \right) a_2 +
    \left( \alpha_1 \beta_2 + T\left(b_1, a_2 \right) \right) a_3  \\
       & - \alpha_1 \, b_2 \times b_3 - \alpha_2 \, b_1 \times b_3
        - \alpha_3  b_1 \times b_2 \\ &- \left\{ a_1, b_2, a_3 \right\} -
        \left\{ a_1, b_3, a_2 \right\} - \left\{a_2, b_1,a_3\right\} \\
    \delta &= -\gamma^{\sigma}  \qquad
      d = -c^{\sigma}  \qquad \text{where} \qquad \sigma = \left(\alpha \leftrightarrow \beta\right)\left(a \leftrightarrow b\right).
\end{split}
\end{equation*}
Here $\sigma$ denotes a permutation of $\alpha$ with $\beta$ and $a$ with $b$, and
\begin{equation}
  \left\{ a,b,c\right\} =  U_{a+c} b - U_a b - U_c b
%
\end{equation}
where $U_a b$ is defined as in \eqref{eq:Uxtimesy}.
\section{Spacetimes over  Jordan algebras of degree three as dilatonic
and spinorial extensions of Min\-ko\-wskian spacetimes\label{sec:SpaceTime-Ext}}

As stated above  we will restrict our studies of generalized
spacetimes to those defined by formally real Jordan algebras of
degree 3.  Our main goal is to give a unified geometric realization
of the conformal and quasiconformal groups of generalized spacetimes
defined by Jordan algebras of degree three and the FTS's defined
over them.

The related geometries in the context of $N=2$ MESGT's were reviewed
in section \ref{sec:review}. The Jordan algebras of degree three
that arose in the study of MESGT's were later studied by Sierra who
showed that there exists a correspondence between them and classical
relativistic point particle actions \cite{Sierra}. In the same work
Sierra showed that this could be extended to a correspondence
between classical relativistic bosonic strings and the Freudenthal
triple systems defined over them.

Consider now the spacetimes coordinatized by the generic Jordan
family
\begin{equation}
     J = \mathbb{R} \oplus \Gamma(Q)
\end{equation}
we shall interpret the extra coordinate corresponding to
$\mathbb{R}$ as a dilatonic coordinate $\rho$ and label the
coordinates defined by $J$ as $(\rho, x_{m}, m=0,1,2,...(d-1) )$ .
The automorphism group $SO(d-1)$ will then be the rotation group of
this space-time under which both the time coordinate $x_0$ and the
dilatonic coordinate $\rho$ will be singlets. The Lorentz group of
this spacetime is the reduced structure group which is simply
\begin{equation}
  \mathrm{SO}(d-1,1) \times \mathrm{SO}(1,1)
\end{equation}
It leaves invariant the cubic norm which, following \cite{Sierra},
we normalize as
\begin{equation}
    \mathcal{V}\left(\rho,x_m \right) = \sqrt{2} \rho \, x_m x_n
    \eta^{mn}
\end{equation}

 Under the action of $SO(d-1,1)$, the dilaton $\rho$
is a singlet and under $SO(1,1)$ we have
\begin{equation}
   SO(1,1): \begin{aligned} \rho &\Rightarrow e^{2\lambda} \rho \\
  x_{m} & \Rightarrow e^{\lambda} x_{m} \end{aligned}
\end{equation}
Freudenthal product of two elements of $J = \mathbb{R} \oplus
\Gamma(Q) $ is simply
\begin{equation}
    (\rho,x) \times (\sigma, y)  = \left(\sqrt{2} x_m y^m , \sqrt{2}\left( \rho y_m +\sigma  x_m \right)
  \right)
\end{equation}
The conformal group of the spacetime is the M\"obius group of $J$ which is
\begin{equation}
   \mathrm{SO}\left(d,2\right) \times \mathrm{SO}\left(2,1\right)
\end{equation}
The Freudenthal triple systems defined over the generic Jordan family
can be represented by $2 \times 2$ matrices
\begin{equation}
   \mathfrak{M}\left( J = \mathbb{R} \oplus \Gamma\left(Q\right)\right) =
  \begin{pmatrix} x^1_d & J^1 \cr J^2 & x^2_d  \end{pmatrix} = X
\end{equation}
where $J^1, J^2 \in J$ and $x^1_d$ and $x^2_d$ are real coordinates.
The automorphism group of $\mathfrak{M}$ is $\mathrm{SO}\left(d,
2\right) \otimes \mathrm{Sp}\left(2, \mathbb{R}\right)$ under which
an element of $\mathfrak{M}$ transforms in the representation
$\left(\mathbf{d+2}, \mathbf{2}\right)$. We shall label the
``coordinates'' of $\mathfrak{M}$  as
\begin{equation*}
   x_\mu^a = \left( x^a_m, x^a_d, \rho^a \right) \quad \text{where} \quad a=1,2
\end{equation*}
and interpret it as coordinates of a conformally covariant phase
space (so that $a=1$ labels the  coordinates  and $a=2$ labels the
momenta).

Skew-symmetric invariant form over $\mathfrak{M}$ is given by
\begin{equation}
   \left<X, Y\right> = \epsilon_{ab} \eta^{\mu\nu} X^a_\mu Y^b_\nu
\end{equation}
We should stress the important fact that  the conformal group of the
spacetime defined by $J$ is isomorphic to the automorphism group of
the Freudenthal triple system $\mathfrak{M}\left(J\right)$ !

To define the quasi-conformal group over the conformal phase space
represented by $\mathfrak{M}\left(J\right)$ we need to extend it by
an extra coordinate corresponding to the cocycle (symplectic form)
over $\mathfrak{M}\left(J\right)$. We shall denote the elements of
$\mathfrak{M}\left(J\right)$ as $X$ and the extra coordinate as $x$.
The quasi-conformal group of $\mathfrak{M}\left(J\right) \oplus
\mathbb{R}$ is the group $\mathrm{SO}\left(d+2, 4\right)$.

The space-times defined by simple Jordan algebras of degree 3
$J_3^\mathbb{A}$ correspond to extensions of Minkowski space-times
in the critical dimensions $d=3,4,6,10$ by a dilatonic ($\rho$)
\emph{and} commuting spinorial coordinates ($\xi^a$).
\begin{equation}
\begin{split}
   J_3^\mathbb{R}  &\Longleftrightarrow \left(\rho, x_m, \xi^{\alpha} \right) \quad m=0,1,2 \quad \alpha =1,2 \\
   J_3^\mathbb{C}  &\Longleftrightarrow \left(\rho, x_m, \xi^{\alpha} \right) \quad m=0,1,2,3 \quad \alpha =1,2,3,4 \\
   J_3^\mathbb{H}  &\Longleftrightarrow \left(\rho, x_m, \xi^{\alpha} \right) \quad m=0,\ldots,5 \quad \alpha =1,\ldots,8 \\
   J_3^\mathbb{C}  &\Longleftrightarrow \left(\rho, x_m, \xi^{\alpha} \right) \quad m=0,\ldots,9 \quad \alpha =1,\ldots,16
\end{split}
\end{equation}
The commuting spinors $\xi$ are represented by a $2 \times 1$ matrix
over $\mathbb{A}=\mathbb{R}, \mathbb{C},\mathbb{H},\mathbb{O}$.  The
cubic norm of a ``vector''  with coordinates $\left(\rho, x_m,
\xi^{\alpha}\right)$ is given by \cite{Sierra}
\begin{equation}
  \mathcal{V}\left(\rho, x_m, \xi^{\alpha}\right) = \sqrt{2} \rho x_m x_n \eta^{mn} + x^m \Bar{\xi} \gamma_m \xi
\end{equation}
The Lorentz groups of the space-times over $J_3^\mathbb{A}$ are
\begin{equation}
   \mathrm{SL}\left(3, \mathbb{R}\right), \quad
    \mathrm{SL}\left(3, \mathbb{C}\right), \quad
    \mathrm{SU}^\ast\left(6\right), \quad \text{and}
    \quad \mathrm{E}_{6(-26)}
\end{equation}
respectively, corresponding to the invariance groups of their cubic
norm. The Freudenthal  product of two vectors in the corresponding
space-time is given by
\begin{multline}
    \left(\rho, x_m, \xi^{\alpha}\right) \times \left(\sigma, y_m, \zeta^{\alpha}\right) = \\
  \left(
     \sqrt{2} \,x_m y^m, \mspace{5mu} 
     \frac{1}{2}\left( \Bar{\xi} \gamma_m \zeta +   
     \Bar{\zeta} \gamma_m \xi \right)
    + \sqrt{2} \left( \rho \, y_m +  \sigma \, x_m \right) , \mspace{5mu}  
     x^m \, \Bar{\zeta} \gamma_m + y^m\, \Bar{\xi}\gamma_m \right)
\end{multline}
The conformal groups of these spacetimes are
\begin{equation}
   \mathrm{Sp}\left(6, \mathbb{R}\right), \quad
   \mathrm{SU}\left(3,3\right), \quad
   \mathrm{SO}^\ast\left(12\right), \quad \text{and} \quad
   \mathrm{E}_{7(-25)}
\end{equation}
respectively. The automorphism groups of the FTS $\mathfrak{M}\left(J_3^\mathbb{A}\right)$ are
isomorphic to their conformal groups.

The quasi-conformal groups acting on
$\mathfrak{M}\left(J_3^\mathbb{A} \oplus \mathbb{R}\right)$, where
$\mathbb{R}$ represents the extra ``cocyle'' coordinate, are
\begin{equation}
  \mathrm{F}_{4(4)}, \quad
  \mathrm{E}_{6(2)}, \quad
  \mathrm{E}_{7(-5)}, \quad \mathrm{E}_{8(-24)}
\end{equation}
whose minimal unitary irreducible representations were constructed
in \cite{GP}.

\section{Geometric realizations of $\mathrm{SO}\left(d+2,4\right)$ as quasiconformal  groups }

Lie algebra of $\mathrm{SO}\left(d+2,4\right)$ admits the following
5-graded decomposition
\begin{equation}
  \mathfrak{so}\left(d+2,4\right) = \mathbf{1} \oplus \left(\mathbf{d+2}, \mathbf{2} \right) \oplus
    \left(\Delta \oplus \mathfrak{sp}\left(2, \mathbb{R}\right) \oplus \mathfrak{so}\left(d,2\right) \right)
   \oplus \left(\mathbf{d+2}, \mathbf{2} \right) \oplus  \mathbf{1}
\end{equation}
Generators are realized as differential operators in $2d+5$
coordinates corresponding to  $\mathfrak{g}^{-2} \oplus
\mathfrak{g}^{-1}$ subspace which we shall denote as $x$ and
$X^{\mu, a}$ where $a=1,2$ is an index of representation
$\mathbf{2}$ of $\mathfrak{sp}\left(2, \mathbb{R}\right)$ and we
shall let the indices $\mu$  run from 1 to $d+2$ with the indices
$d+1$  and $d+2$  labelling the two  timelike coordinates, i.e
$x_{\mu}$ transforms like a vector of $SO(d,2)$.

Let $\epsilon_{ab}$ be symplectic real-valued matrix, and
$\eta_{\mu\nu}$ denote signature $\left(d,2\right)$ metric preserved
by $SO(d,2)$. Then
\begin{equation}
   \mathcal{I}_4 = \eta_{\mu\nu} \eta_{\rho\tau} \epsilon_{ac}  \epsilon_{bd} X^{\mu, a} X^{\nu, b} X^{\rho, c} X^{\tau, d}
\end{equation}
is  a 4th-order polynomial invariant under the semisimple part of
$\mathfrak{g}^0$. Define
\begin{equation}
\begin{split}
   K_+ &= \frac{1}{2} \left( 2 x^2 - \mathcal{I}_4 \right)  \frac{\partial}{\partial x} -
  \frac{1}{4} \frac{\partial \mathcal{I}_4}{ \partial X^{\mu, a} } \eta^{\mu\nu} \epsilon^{ab} \frac{\partial}{\partial X^{\nu, b}} +  x \, X^{\mu, a} \frac{\partial}{\partial X^{\mu, a}} \\
  U_{\mu, a} &= \frac{\partial}{\partial X^{\mu, a}} - \eta_{\mu, \nu} \epsilon_{a b} X^{\nu, b} \frac{\partial}{\partial x} \\
  M_{\mu\nu} &= \eta_{\mu\rho} X^{\rho,a} \frac{\partial}{\partial X^{\nu,a}} -
                \eta_{\nu\rho} X^{\rho,a} \frac{\partial}{\partial X^{\mu,a}} \\
  J_{ab} &= \epsilon_{ac} X^{\mu, c}  \frac{\partial}{\partial X^{\mu,b}} +
            \epsilon_{bc} X^{\mu, c}  \frac{\partial}{\partial X^{\mu,a}} \\
   K_- &= \frac{\partial}{\partial x}  \phantom{ and also}
  \Delta = 2 x \frac{\partial}{\partial x} + X^{\mu, a} \frac{\partial}{\partial X^{\mu, a}}
   \phantom{ and also}   \Tilde{U}_{\mu, a} = \left[ U_{\mu, a}, K_+ \right]
\end{split}
\end{equation}
where $\epsilon^{ab}$ denotes an inverse symplectic metric:
$\epsilon^{ab} \epsilon_{bc} = {\delta^a}_c$ and $\Tilde{U}_{\mu,
a}$ evaluates to
\begin{equation}
\begin{split}
    \Tilde{U}_{\mu, a} &=   \eta_{\mu\nu} \epsilon_{ad} \left( \eta_{\lambda\rho} \epsilon_{bc}
                           X^{\nu,b} X^{\lambda, c} X^{\rho, d} - x X^{\nu, d} \right) \frac{\partial}{\partial x}
         +  x \frac{\partial}{\partial X^{\mu, a}} \\ &-
           \eta_{\mu\nu}\epsilon_{ab} X^{\nu,b} X^{\rho, c}\frac{\partial}{\partial X^{\rho, c}}
         -  \epsilon_{ad} \eta_{\lambda \rho} X^{\rho, d} X^{\lambda, c} \frac{\partial}{\partial X^{\mu, c}}
         \\ &+  \epsilon_{ad} \eta_{\mu\nu} X^{\rho, d} X^{\nu, b} \frac{\partial}{\partial X^{\rho, b}}
          +  \eta_{\mu\nu} \epsilon_{bc} X^{\nu, b} X^{\rho, c} \frac{\partial}{\partial X^{\rho, a}}
\end{split}
\end{equation}
we have
\begin{equation*}
   \frac{\partial \mathcal{I}_4}{ \partial X^{\mu, a} } = - 4 \, \eta_{\mu\nu} \, \eta_{\lambda\rho} \,
     X^{\nu,b} X^{\lambda, c}  X^{\rho, d} \, \epsilon_{bc} \epsilon_{ad}
\end{equation*}
These generators satisfy the following commutation relation:
\begin{subequations}
\label{eq:sod4algebra}
\begin{equation}
\begin{split}
    \left[ M_{\mu\nu}, M_{\rho\tau} \right] &= \eta_{\nu\rho} M_{\mu\tau} - \eta_{\mu\rho} M_{\nu\tau} + \eta_{\mu\tau} M_{\nu\rho} - \eta_{\nu\tau} M_{\mu\rho} \\
    \left[ J_{ab} , J_{cd} \right] &= \epsilon_{cb} J_{ad} + \epsilon_{ca} J_{bd} + \epsilon_{db} J_{ac} + \epsilon_{da} J_{bc} \\
   \left[ \Delta, K_\pm \right] &= \pm 2 K_\pm \phantom{also} \left[ K_-, K_+ \right] = \Delta \\
   \left[ \Delta, U_{\mu, a} \right] &= - U_{\mu, a} \phantom{also}
   \left[ \Delta, \Tilde{U}_{\mu, a} \right] =  \Tilde{U}_{\mu, a} \\
   \left[ U_{\mu, a} , K_+ \right] &= \Tilde{U}_{\mu, a} \phantom{also }
   \left[ \Tilde{U}_{\mu, a} , K_- \right] = -U_{\mu, a} \\
    \left[ U_{\mu, a}, U_{\nu, b} \right] &= 2 \eta_{\mu\nu} \epsilon_{ab} K_- \phantom{also}
    \left[ \Tilde{U}_{\mu, a}, \Tilde{U}_{\nu, b} \right] = 2 \eta_{\mu\nu} \epsilon_{ab} K_+
\end{split}
\end{equation}
\begin{equation}
\begin{split}
   \left[ M_{\mu\nu}, U_{\rho, a} \right] &= \eta_{\nu\rho} U_{\mu, a} - \eta_{\mu\rho} U_{\nu, a}  \phantom{also}
   \left[ M_{\mu\nu}, \Tilde{U}_{\rho, a} \right] = \eta_{\nu\rho} \Tilde{U}_{\mu, a} - \eta_{\mu\rho} \Tilde{U}_{\nu, a} \\
   \left[ J_{ab}, U_{\mu, c} \right] &= \epsilon_{cb} U_{\mu,a} + \epsilon_{ca} U_{\mu, b}  \phantom{also}
   \left[ J_{ab}, \Tilde{U}_{\mu, c} \right] = \epsilon_{cb} \Tilde{U}_{\mu,a} + \epsilon_{ca} \Tilde{U}_{\mu, b} \\
\end{split}
\end{equation}
\begin{equation}
  \left[ U_{\mu, a} , \Tilde{U}_{\nu, b} \right] = \eta_{\mu\nu} \epsilon_{ab} \, \Delta - 2 \, \epsilon_{ab} M_{\mu\nu} -
          \eta_{\mu\nu} J_{ab}
\end{equation}
\end{subequations}
The distance invariant under $\mathrm{SO}\left(d+2,4\right)$ can be
constructed following \cite{GKN1}. Let us first introduce a
difference between two vectors $\mathcal{X}$ and $\mathcal{Y}$ on
$\mathfrak{g}^{-1} \oplus \mathfrak{g}^{-2}$:
\begin{equation}
    \delta \left( \mathcal{X}, \mathcal{Y} \right) = \left( X^{\mu, a} - Y^{\mu, a}, x-y - \eta_{\mu\nu}\epsilon_{ab} X^{\mu,a} Y^{\nu,b} \right)
\end{equation}
 and define the ``length'' of a vector $\mathcal{X}$ as
\begin{equation}
   \ell \left(\mathcal{X} \right) = \mathcal{I}_4\left(X\right) + 2 x^2
\end{equation}
Then the  cone defined by $\ell\left( \delta\left(\mathcal{X},
\mathcal{Y}\right)\right) =0$ is invariant w.r.t. the full group
$SO(d+2,4)$, because of the following identities:
\begin{equation}
\begin{split}
   \Delta \circ \ell\left( \delta\left(\mathcal{X}, \mathcal{Y}\right)\right) &=
        4 \, \ell\left( \delta\left(\mathcal{X}, \mathcal{Y}\right)\right) \\
   \Tilde{U}_{\mu, a} \circ \ell\left( \delta\left(\mathcal{X}, \mathcal{Y}\right)\right) &=
         -2 \eta_{\mu\nu}\epsilon_{ab} \left( X^{\nu, b} + Y^{\nu,b} \right)
          \ell\left( \delta\left(\mathcal{X}, \mathcal{Y}\right)\right) \\
    K_+ \circ \ell\left( \delta\left(\mathcal{X}, \mathcal{Y}\right)\right) &=
         2  \left( x + y \right)
          \ell\left( \delta\left(\mathcal{X}, \mathcal{Y}\right)\right) \\
    \text{any other generator} \circ \ell\left( \delta\left(\mathcal{X}, \mathcal{Y}\right)\right) &=  0
\end{split}
\end{equation}

\section{Geometric realizations of $\mathrm{E}_{8(-24)},
\mathrm{E}_{7(-5)}, \mathrm{E}_{6(2)}$ and $\mathrm{F}_{4(4)}$ as
quasiconformal groups}

The minimal unitary representations of the quasiconformal groups of
the spacetimes defined by simple formally real Jordan algebras of
degree three were given in our earlier paper \cite{GP}. In this
section we will give their geometric realizations as quasiconformal
groups in an $SO(d,2) \times Sp(2,\mathbb{R}) $ covariant basis
where $d$ is equal to one of the critical dimensions $3,4,6,10$.

\subsection{Geometric realization of the quasiconformal group $\mathrm{E}_{8(-24)}$ }

For realizing  the geometric action of the quasiconformal group
$\mathrm{E}_{8(-24)}$ in an $\mathrm{SO}(10,2) \times
\mathrm{SO}(2,1)$ covariant basis we shall use the following
5-graded decomposition of its Lie algebra
\begin{equation}
 \mathfrak{e}_{8(-24)} = \tilde{\mathbf{1}} \oplus \tilde{\mathbf{56}} \oplus
  \left[  \mathfrak{so}(1,1) \oplus \mathfrak{e}_{7(-25)}  \right] \oplus \mathbf{56}
  \oplus \mathbf{1}
\end{equation}
\begin{multline*}
    \mathfrak{e}_{8(-24)} = \mathbf{1} \oplus \left(
      \begin{aligned}  \left(\mathbf{2}, \mathbf{12}\right)\,\, \\
       \left(\mathbf{1}, \mathbf{32}_c \right) \end{aligned} \right)
     \\ \oplus
   \left[ \Delta
   \oplus \left( \begin{aligned}
        & \phantom{also11}\left( \mathbf{2} , \mathbf{32}_s \right) \cr
        & \mathfrak{sp}\left(2, \mathbb{R} \right) \oplus
              \mathfrak{so}\left(10,2\right)
           \end{aligned} \right)
     \right]
  \oplus \left(   \begin{aligned}  \left(\mathbf{2}, \mathbf{12}\right)\,\, \\
        \left(\mathbf{1}, \mathbf{32}_c\right) \end{aligned}
   \right) \oplus \mathbf{1}.
\end{multline*}
The generators of the simple subalgebra $\mathfrak{e}_{7(-25)}$ in
$\mathfrak{g}^0$ satisfy the following  $\mathrm{SO}(10,2)$
covariant commutation relations.
\begin{equation}
\label{eq:e7linear}
\begin{split}
     \left[ M_{\mu\nu}, Q_{a \Dot{\alpha} } \right] &=
   Q_{a \Dot{\beta}} \, {\left( \Gamma_{\mu\nu} \right)^{\Dot{\beta}}}_{\Dot{\alpha}} \\
     \left[ J_{ab}, Q_{c \Dot{\alpha}} \right] &= \epsilon_{cb} Q_{a \Dot{\alpha}} +
          \epsilon_{ca} Q_{b \Dot{\alpha}} \\
   \left[ Q_{a \Dot{\alpha}} ,  Q_{b \Dot{\beta}} \right] &=
            \epsilon_{ab} \left( C \Gamma_{\mu\nu}\right)_{\Dot{\alpha} \Dot{\beta} } M^{\mu\nu} +
         C_{\Dot{\alpha}\Dot{\beta}} J_{ab}
\end{split}
\end{equation}
where $M_{\mu\nu}$ and $J_{ab}$ are the generators of
$\mathrm{SO}(10,2)$ and $\mathrm{Sp}(2,\mathbb{R})$, respectively
and $Q_{a \Dot{\alpha}}$  are the remaining generators transforming
in the $(\mathbf{32}_s, \mathbf{2}))$ of $\mathrm{SO}(10,2) \times
Sp(2,\mathbb{R})$ . $C$ is the charge conjugation matrix in $(10,2)$
dimensions and is antisymmetric
\begin{equation}
  C^t=-C
\end{equation}
The generators of $\mathrm{E}_{7(-25)}$ are realized in terms of the
``coordinates'' $X^{\mu,a}$ and $\psi^\alpha$ transforming in the
$(\mathbf{12},\mathbf{2})$ and $(\mathbf{32}_c,\mathbf{1})$
representation of $\mathrm{SO}(10,2) \times Sp(2,\mathbb{R})$ as
follows
\begin{equation}
\begin{split}
     M_{\mu\nu} &= \eta_{\mu\rho} X^{\rho,a} \frac{\partial}{\partial X^{\nu,a}} -
                \eta_{\nu\rho} X^{\rho,a} \frac{\partial}{\partial X^{\mu,a}} -
           \psi^\alpha { \left(\Gamma_{\mu\nu} \right)^\beta}_{\alpha} \frac{\partial}{\partial \psi^\beta} \\
    J_{ab} &= \epsilon_{ac} X^{\mu, c}  \frac{\partial}{\partial X^{\mu,b}} +
            \epsilon_{bc} X^{\mu, c}  \frac{\partial}{\partial X^{\mu,a}} \\
    Q_{a\Dot{\alpha}} &= \epsilon_{ab} X^{\mu, b} {\left(\Gamma_\mu \right)^\beta}_{\Dot{\alpha}}
  \frac{\partial}{\partial \psi^\beta} -  \psi^\beta
   \left(C \Gamma_\mu\right)_{\beta \Dot{\alpha}} \eta^{\mu\nu} \frac{\partial}{\partial X^{\nu, a}}
\end{split}
\end{equation}
where $\Gamma_{\mu}$ are the gamma matrices and $\Gamma_{\mu\nu} =
\frac{1}{4} \left( \Gamma_\mu \Gamma_\nu - \Gamma_\nu
\Gamma_\mu\right)$.  $\alpha, \beta,..$ and $\Dot{\alpha},
\Dot{\beta}, ...$ are chiral and anti-chiral spinor indices that run
from 1 to 32, respectively. $\Gamma$ matrices are taken to be in a
chiral basis (with $\Gamma_{13}$ being diagonal) . The spinorial
``coordinates'' $\psi^{\alpha}$ transform as a Majorana-Weyl spinor of
$\mathrm{SO}(10,2)$.  One convenient choice for gamma matrices is
\begin{equation}
\begin{split}
   &\Gamma_{i} = \sigma_1 \otimes \sigma_1 \otimes \Gamma_i^{(8)} \quad
   \Gamma_9 = \sigma_1 \otimes \sigma_1 \otimes  \Gamma_9^{(8)} \quad
   \Gamma_{10} = \sigma_1 \otimes \sigma_3 \otimes 1_{16} \quad
   \\ & \Gamma_{11} = \sigma_1 \otimes i \sigma_2 \otimes 1_{16} \quad
   \Gamma_{12} = i \sigma_2 \otimes 1_{32} \quad
   C = 1_2 \otimes i \sigma_2 \otimes 1_{16}
\end{split}
\label{eq:clifalg}
\end{equation}
where $i=1,\ldots,8$ and $\Gamma_9^{(8)} = \Gamma_1^{(8)} \dots \Gamma_8^{(8)}$. Matrices
$\Gamma_i^{(8)}$ are those of Clifford algebra of $\mathbb{R}^8$. The chiral realization
\eqref{eq:clifalg} assumes mostly plus signature convention:
\begin{equation}
     \eta_{\mu\nu} = \mathrm{diag} \left( \left(+\right)^{10}, \left(-\right)^2\right) \qquad
      \mu,\nu=1,\ldots,12.
\end{equation}
The fourth order invariant of $\mathfrak{e}_{7(-25)}$ in the above basis reads as
\begin{equation}
\begin{split}
   \mathcal{I}_4 &= \eta_{\mu\nu} \eta_{\rho\tau} \epsilon_{ac}  \epsilon_{bd} X^{\mu, a} X^{\nu, b} X^{\rho, c} X^{\tau, d}
     + 2 \epsilon_{ab} X^{\mu, a} X^{\nu, b} \psi^\alpha \left(C \Gamma_{\mu\nu} \right)_{\alpha\beta} \psi^\beta \\
     &+ \frac{1}{6} \psi^\alpha \left(C \Gamma_{\mu\nu} \right)_{\alpha\beta} \psi^\beta
       \psi^\gamma \left(C \Gamma^{\mu\nu} \right)_{\gamma\delta} \psi^\delta
\end{split}
\end{equation}
Given the above data, it is straightforward to realize the
generators of $\mathfrak{e}_{8(-24)}$ on a $56+1 =57$ dimensional
space following \cite{GKN1,GKN2,GP}. We start with negative grade
generators
\begin{equation}
\begin{split}
    K_- &= \frac{\partial}{\partial x} \qquad
    U_{\alpha} = \frac{\partial}{\partial \psi^\alpha} - C_{\alpha\beta} \psi^\beta \frac{\partial}{\partial x} \\
    U_{\mu, a} &= \frac{\partial}{\partial X^{\mu, a}} - \eta_{\mu, \nu} \epsilon_{a b} X^{\nu, b} \frac{\partial}{\partial x}
\end{split}
\end{equation}
where $x$ is the singlet ``cocycle'' coordinate. Grade +2 generator is
\begin{equation}
\begin{split}
K_+ &= \frac{1}{2} \left( 2 x^2 - \mathcal{I}_4 \right)  \frac{\partial}{\partial x} -
  \frac{1}{4} \frac{\partial \mathcal{I}_4}{ \partial X^{\mu, a} }
  \eta^{\mu\nu} \epsilon^{ab} \frac{\partial}{\partial X^{\nu, b}} \\ &+
  \frac{1}{4} \frac{\partial \mathcal{I}_4}{ \partial \psi^\alpha } \left(C^{-1}\right)^{\alpha\beta}
     \frac{\partial}{\partial \psi^\beta}
  +  x \, X^{\mu, a} \frac{\partial}{\partial X^{\mu, a}}
  +  x \, \psi^\alpha \frac{\partial}{\partial \psi^\alpha}
\end{split}
\end{equation}
Generators of grade +1 space are obtained by commuting $K_+$ with corresponding generators of $\mathfrak{g}^{-1}$:
\begin{equation}
    \Tilde{U}_{\mu, a} = \left[ U_{\mu, a}, K_+ \right] \qquad
    \Tilde{U}_{\alpha} = \left[ U_{\alpha}, K_+ \right] .
\end{equation}
The generator that determines the five grading is simply
\begin{equation}
   \Delta = 2 x \frac{\partial}{\partial x} + X^{\mu, a} \frac{\partial}{\partial X^{\mu, a}} +
   \psi^\alpha \frac{\partial}{\partial \psi^\alpha}
\end{equation}
The commutation relations of these generators are those of
\eqref{eq:sod4algebra} for $d=10$  supplemented with
\eqref{eq:e7linear} and the following:
\begin{equation}
\begin{split}
  &\begin{aligned}
     \left[ U_\alpha, U_{\beta} \right] &= 2 \, C_{\alpha\beta} K_- \\
     \left[ \Tilde{U}_\alpha, \Tilde{U}_{\beta} \right] &= 2 \, C_{\alpha\beta} K_+
  \end{aligned}
   \qquad\qquad
  \begin{aligned}
    \left[ U_{\alpha}, K_+ \right] &= \Tilde{U}_{\alpha} \\
    \left[ \Tilde{U}_{\alpha}, K_- \right] &= - U_{\alpha}
  \end{aligned} \\
 &\begin{aligned}
    \left[ Q_{a \Dot{\alpha}}, U_{\mu, b} \right] &= -\epsilon_{ab} {\left( \Gamma_{\mu} \right)^\alpha}_{\Dot{\alpha}}
            U_{\alpha} \\
    \left[ Q_{a \Dot{\alpha}}, \Tilde{U}_{\mu, b} \right] &= -\epsilon_{ab} {\left( \Gamma_{\mu} \right)^\alpha}_{\Dot{\alpha}}
            \Tilde{U}_{\alpha}
 \end{aligned}
  \qquad
   \begin{aligned}
   \left[ Q_{a \Dot{\alpha}}, U_{\beta} \right] &= \left( C\Gamma_\mu \right)_{\beta\Dot{\alpha}} \eta^{\mu\nu} U_{\nu, a} \\
   \left[ Q_{a \Dot{\alpha}}, \Tilde{U}_{\beta} \right] &=
       \left( C\Gamma_\mu \right)_{\beta\Dot{\alpha}} \eta^{\mu\nu} \Tilde{U}_{\nu, a}
   \end{aligned}
 \\
 & \begin{aligned}
        \left[ \Tilde{U}_{\alpha}, U_{\mu, a} \right] &=
           -{\left( C \Gamma_\mu C^{-1} \right)_{\alpha}}^{\Dot{\alpha}} Q_{a \Dot{\alpha}} \\
       \left[ U_{\alpha}, \Tilde{U}_{\mu, a} \right] &=
                {\left( C \Gamma_\mu C^{-1} \right)_{\alpha}}^{\Dot{\alpha}} Q_{a \Dot{\alpha}}
   \end{aligned}
   \quad
    \left[ U_{\alpha}, \Tilde{U}_{\beta} \right] = C_{\alpha\beta} \Delta-\left(C \Gamma_{\mu\nu} \right)_{\alpha\beta} M^{\mu\nu}
\end{split}
\end{equation}
with all  the remaining commutators vanishing. The explicit
expressions for the grade $+1$ generators are
\begin{equation}
 \begin{split}
    \Tilde{U}_{\mu, a} &= -\frac{1}{4} \frac{\partial \mathcal{I}_4}{\partial X^{\mu, a}} \frac{\partial}{\partial x} -
            x \, \eta_{\mu\nu} \epsilon_{ab} X^{\nu,b} \frac{\partial}{\partial x} +
            x \, \frac{\partial}{\partial X^{\mu, a}} \\
   &- \frac{1}{4} \frac{\partial^2 \mathcal{I}_4}{\partial X^{\mu, a} \partial X^{\nu, b}}
     \eta^{\nu\lambda} \epsilon^{bc} \frac{\partial}{\partial X^{\lambda,c}} -
     \frac{1}{4}  \frac{\partial^2 \mathcal{I}_4}{\partial X^{\mu, a} \partial \psi^\alpha} \left(C^{-1}\right)^{\alpha\beta}
       \frac{\partial}{\partial \psi^\beta} \\
        & - \eta_{\mu\nu} \epsilon_{ab} X^{\nu,b} \left( X^{\lambda,c}
        \frac{\partial}{\partial X^{\lambda,c}} +  \psi^\gamma \frac{\partial}{\partial \psi^\gamma} \right)
 \end{split}
\end{equation}
\begin{equation}
\begin{split}
    \Tilde{U}_\alpha &= -\frac{1}{4} \frac{\partial \mathcal{I}_4}{\partial \psi^\alpha} \frac{\partial}{\partial x} -
            x \left( C_{\alpha \beta} \psi^\beta \right) \frac{\partial}{\partial x} - C_{\alpha\beta} \psi^\beta \left( X^{\mu, a} \frac{\partial}{\partial X^{\mu, a}} +
           \psi^\gamma \frac{\partial}{\partial \psi^\gamma} \right)
            \\ &- \frac{1}{4} \frac{\partial^2
               \mathcal{I}_4}{\partial X^{\mu, a} \partial \psi^\alpha} \eta^{\mu\nu} \epsilon^{ab}
                 \frac{\partial}{\partial X^{\nu,b}} -
            \frac{1}{4} \frac{\partial^2 \mathcal{I}_4}{\partial \psi^\alpha \partial \psi^\beta}
                 \left(C^{-1}\right)^{\beta\gamma} \frac{\partial}{\partial \psi^\gamma} +
           x \frac{\partial}{\partial \psi^\alpha}
\end{split}
\end{equation}
The above geometric realization of the quasiconformal action of the
Lie algebra of  $\mathrm{E}_{8(-24)}$ can be consistently truncated
to the quasiconformal realizations of $\mathrm{E}_{7(-5)}$,
$\mathrm{E}_{6(2)}$ and $\mathrm{F}_{4(4)}$, which we discuss in the
following subsections. We should stress that for all these groups
one can define a quartic norm such that they leave the generalized
light-cone defined with respect to this quartic norm invariant as
was shown for the maximally split exceptional groups in \cite{GKN1}
and for $SO(d+2,4)$ in section 5 above.

\subsection{Geometric realization of the quasiconformal group $\mathrm{E}_{7(-5)}$}

Truncation of the geometric realization of the quasiconformal group
$\mathrm{E}_{8(-24)}$ to $\mathfrak{e}_{7(-5)}$ is achieved by
``dimensional reduction'' from 10 to 6 dimensions. Reduction of
32-component Majorana-Weyl spinor of
$\mathfrak{so}\left(10,2\right)$ is done by using the projection
operators:
\begin{equation}
    {\mathcal{P}^\alpha}_\beta = \frac{1}{2} {\left( 1 +  \Gamma_1\Gamma_2 \Gamma_3 \Gamma_4 \right)^\alpha}_\beta
   \qquad
   {\mathcal{P}^{\Dot{\alpha}}}_{\Dot{\beta}} = \frac{1}{2} {\left( 1 +  \Gamma_1 \Gamma_2 \Gamma_3 \Gamma_4
   \right)^{\Dot{\alpha}}}_{\Dot{\beta}}
\end{equation}
where we assumed we compactify first 4 compact directions. This projection will reduce number of spinor components
down to 16. It is clear that projected spinors will have the same chirality as their ancestors:
\begin{equation}
      \mathcal{P}  \Gamma_{5}\ldots \Gamma_{12}  \mathcal{P}  =  \mathcal{P} \Gamma_{13}  \mathcal{P}
\end{equation}
This 16-component spinor would thus comprise 2 same chirality
8-\-compo\-nents spinors of $\mathfrak{so}\left(6,2\right)$
satisfying symplectic Majorana-Weyl reality condition. Their R-group
is $\mathfrak{su}\left(2\right)$ - part of the
$\mathfrak{so}\left(4\right)$ of the transverse directions that
leaves the projection operator invariant. Thus the relevant
5-graded decomposition of $\mathfrak{e}_{7(-5)}$
\begin{equation*}
\mathfrak{e}_{7(-5)} = \Tilde{\mathbf{1}} \oplus \Tilde{\mathbf{32}} \oplus [
\mathfrak{so}^\ast(12) \oplus \mathfrak{so}(1,1) ] \oplus \mathbf{32} \oplus \mathbf{1}
\end{equation*}
reads
\begin{equation}
\begin{split}
  \mathfrak{e}_{7(-5)} &= \mathbf{1} \oplus
    \left(\begin{aligned} \left(\mathbf{2}, \mathbf{1}, \mathbf{8}_v \right) \\
                          \left(\mathbf{1}, \mathbf{2}, \mathbf{8}_c \right)
          \end{aligned}\right) \oplus \\
     &\left[ \Delta \oplus \left[
         \begin{aligned} & \left( \mathbf{2}, \mathbf{2}, \mathbf{8}_s\right) \cr
          \mathfrak{sp}\left(2, \mathbb{R}\right) &\oplus \mathfrak{su}\left(2\right) \oplus \mathfrak{so}\left(6,2\right)
    \end{aligned} \right]  \right] \oplus
    \left(\begin{aligned} \left(\mathbf{2}, \mathbf{1}, \mathbf{8}_v \right) \\
                          \left(\mathbf{1}, \mathbf{2}, \mathbf{8}_c \right)
      \end{aligned}\right) \oplus
    \mathbf{1}
\end{split}
\end{equation}
Let $\xi^{i,\alpha}$ be an $\mathfrak{su}\left(2\right)$ doublet  of
$\mathfrak{so}\left(6,2\right)$ chiral spinors (symplectic
Majorana-Weyl spinor) with $a,b,..=1,2$ and $\alpha,
\beta,...=1,2,..,8$. Then one can realize the Lie algebra of
$\mathfrak{so}^\ast(12)$ of grade zero subspace as
\begin{equation}
\begin{split}
     M_{\mu\nu} &= \eta_{\mu\rho} X^{\rho,a} \frac{\partial}{\partial X^{\nu,a}} -
                \eta_{\nu\rho} X^{\rho,a} \frac{\partial}{\partial X^{\mu,a}} -
           \xi^{i,\alpha} { \left(\Gamma_{\mu\nu} \right)^\beta}_{\alpha} \frac{\partial}{\partial \xi^{i,\beta}} \\
    J_{ab} &= \epsilon_{ac} X^{\mu, c}  \frac{\partial}{\partial X^{\mu,b}} +
            \epsilon_{bc} X^{\mu, c}  \frac{\partial}{\partial X^{\mu,a}} \\
    L_{ij} &= \epsilon_{ik} \xi^{k, \alpha} \frac{\partial}{\partial \xi^{j, \alpha}} +
              \epsilon_{jk} \xi^{k, \alpha} \frac{\partial}{\partial \xi^{i, \alpha}} \\
    Q_{i a\Dot{\alpha}} &= \epsilon_{ab} X^{\mu, b} {\left(\Gamma_\mu \right)^\beta}_{\Dot{\alpha}}
  \frac{\partial}{\partial \xi^{i, \beta}} -  \epsilon_{ij} \psi^{j,\beta}
   \left(C \Gamma_\mu\right)_{\beta \Dot{\alpha}} \eta^{\mu\nu} \frac{\partial}{\partial X^{\nu, a}}
\end{split}
\end{equation}
where $C^t=C$ and $\mu,\nu,.. =1,2,..,8$.  Generators $M, J, L, Q$
form $\mathfrak{so}^\ast\left(12\right)$ algebra:
\begin{equation}
\label{eq:d6linear}
\begin{split}
     \left[ M_{\mu\nu}, Q_{i a \Dot{\alpha} } \right] &=
   Q_{i a \Dot{\beta}} \, {\left( \Gamma_{\mu\nu} \right)^{\Dot{\beta}}}_{\Dot{\alpha}} \\
     \left[ L_{ij}, L_{km} \right] &= \epsilon_{kj} L_{im} + \epsilon_{ki} L_{jm} + \epsilon_{mj} L_{ik} +
            \epsilon_{mi} L_{jk}   \\
     \left[ J_{ab}, Q_{i c \Dot{\alpha}} \right] &= \epsilon_{cb} Q_{i a \Dot{\alpha}} +
          \epsilon_{ca} Q_{i b \Dot{\alpha}} \\
   \left[ Q_{i a \Dot{\alpha}} ,  Q_{j b \Dot{\beta}} \right] &=
            \epsilon_{ij} \epsilon_{ab} \left( C \Gamma_{\mu\nu}\right)_{\Dot{\alpha} \Dot{\beta} } M^{\mu\nu} +
         \epsilon_{ij} C_{\Dot{\alpha}\Dot{\beta}} J_{ab} + \epsilon_{ab} C_{\Dot{\alpha}\Dot{\beta}} L_{ij}
\end{split}
\end{equation}
corresponding to the decomposition
\begin{equation}
\mathfrak{so}^\ast\left(12\right) \supset \mathfrak{so}^\ast\left(8\right) \oplus
\mathfrak{so}^\ast\left(4\right) \equiv \mathfrak{so}\left(6,2\right) \oplus \mathfrak{su}\left(2\right)
\oplus \mathfrak{sp}\left(2, \mathbb{R}\right)
\end{equation}
The fourth order invariant of $\mathfrak{so}^\ast\left(12\right)$ in the above basis is given by
\begin{equation}
\begin{split}
   \mathcal{I}_4 &= \eta_{\mu\nu} \eta_{\rho\tau} \epsilon_{ac}  \epsilon_{bd} X^{\mu, a} X^{\nu, b} X^{\rho, c} X^{\tau, d}
     - 2 \epsilon_{ij} \epsilon_{ab} X^{\mu, a} X^{\nu, b} \xi^{i,\alpha}
           \left(C \Gamma_{\mu\nu} \right)_{\alpha\beta} \xi^{j,\beta} \\
     &+ \frac{1}{4} \xi^{i, \alpha} \left(C \Gamma_{\mu\nu} \right)_{\alpha\beta} \xi^{j,\beta}
       \xi^{k,\gamma} \left(C \Gamma^{\mu\nu} \right)_{\gamma\delta} \xi^{l,\delta} \epsilon_{ij} \epsilon_{kl}
\end{split}
\end{equation}
We can now write generators of $\mathfrak{e}_{7(-5)}$, starting with
negative grade generators
\begin{equation}
\begin{split}
    K_- &= \frac{\partial}{\partial x} \qquad
    U_{i, \alpha} = \frac{\partial}{\partial \xi^{i,\alpha}} +
             \epsilon_{ij} C_{\alpha\beta} \xi^{j,\beta} \frac{\partial}{\partial x} \\
    U_{\mu, a} &= \frac{\partial}{\partial X^{\mu, a}} - \eta_{\mu, \nu} \epsilon_{a b} X^{\nu, b} \frac{\partial}{\partial x}
\end{split}
\end{equation}
Positive grade $+2$ generator $K_+$ is
\begin{equation}
\begin{split}
K_+ &= \frac{1}{2} \left( 2 x^2 - \mathcal{I}_4 \right)  \frac{\partial}{\partial x} -
  \frac{1}{4} \frac{\partial \mathcal{I}_4}{ \partial X^{\mu, a} }
  \eta^{\mu\nu} \epsilon^{ab} \frac{\partial}{\partial X^{\nu, b}} \\ &+
  \frac{1}{4} \epsilon^{ij} \frac{\partial \mathcal{I}_4}{ \partial \xi^{i,\alpha} } \left(C^{-1}\right)^{\alpha\beta}
     \frac{\partial}{\partial \xi^{j,\beta}}
  +  x \, X^{\mu, a} \frac{\partial}{\partial X^{\mu, a}}
  +  x \, \xi^{i,\alpha} \frac{\partial}{\partial \xi^{i,\alpha}}
\end{split}
\end{equation}
Commutation relations of $\mathfrak{g}^{-1}$ and $\mathfrak{g}^{+1}$
specific to $6+2=8$ dimensions are :
\begin{equation}
\begin{split}
   \left[ U_{i, \alpha}, \Tilde{U}_{j, \beta} \right] &= \epsilon_{ij} \left(C \Gamma_{\mu\nu}\right)_{\alpha\beta} M^{\mu\nu}
     + C_{\alpha\beta} L_{ij}  - \epsilon_{ij} C_{\alpha\beta} \Delta \\
   \left[ U_{i, \alpha}, \Tilde{U}_{\mu, a} \right] &= -
        {\left( C \Gamma_\mu C^{-1} \right)_\alpha}^{\Dot{\alpha}} Q_{i a \Dot{\alpha}}.
\end{split}
\end{equation}
Grade $+1$ generators have the following form:
\begin{equation}
\begin{split}
   \Tilde{U}_{\mu,a} &= - \frac{1}{4} \frac{\partial \mathcal{I}_4}{\partial X^{\mu,a}} \frac{\partial}{\partial y} -
      \eta_{\mu\nu}\epsilon_{ab} X^{\nu, b} \,y \frac{\partial}{\partial y} + y \frac{\partial}{\partial X^{\mu, a}} \\
    & -\frac{1}{4}\frac{\partial^2 \mathcal{I}_4}{\partial X^{\mu, a} \partial X^{\nu, b}} \eta^{\nu \rho}\epsilon^{bc}
     \frac{\partial}{\partial X^{\rho, c}} +
    \frac{1}{4} \frac{\partial^2 \mathcal{I}_4}{\partial X^{\mu, a} \xi^{i, \alpha}} \epsilon^{ij}
    \left(C^{-1}\right)^{\alpha\beta} \frac{\partial}{\partial \xi^{j, \beta}}\\
   & - \eta_{\mu\nu}\epsilon_{ab} X^{\nu,b} \left( X^{\lambda,c}\frac{\partial}{\partial X^{\lambda,c}} +
        \xi^{i,\alpha} \frac{\partial}{\partial \xi^{i, \alpha}} \right)
\end{split}
\end{equation}
\begin{equation}
\begin{split}
   \Tilde{U}_{i,\alpha} &= - \frac{1}{4} \frac{\partial \mathcal{I}_4}{\partial \xi^{i,\alpha}} \frac{\partial}{\partial y} +
      C_{\alpha\beta}\epsilon_{ij} \xi^{j, \beta} \,y \frac{\partial}{\partial y} +
       y \frac{\partial}{\partial \xi^{i, \alpha}} \\
    & +\frac{1}{4}\frac{\partial^2 \mathcal{I}_4}{\partial \xi^{i, \alpha} \partial \xi^{j, \beta}}
     \left(C^{-1}\right)^{\beta \gamma}\epsilon^{jk}
     \frac{\partial}{\partial \xi^{k, \gamma}} -
    \frac{1}{4} \frac{\partial^2 \mathcal{I}_4}{\partial \xi^{i, \alpha} X^{\mu, a}} \eta^{\mu\nu} \epsilon^{ab}
     \frac{\partial}{\partial X^{\nu, b}}\\
   & +C_{\alpha\beta}\epsilon_{ij} \xi^{j,\beta} \left( X^{\lambda,c}\frac{\partial}{\partial X^{\lambda,c}} +
        \xi^{i,\alpha} \frac{\partial}{\partial \xi^{i, \alpha}} \right)
\end{split}
\end{equation}

\subsection{Geometric realization of the quasiconformal group $\mathrm{E}_{6(2)}$}

Truncation to $\mathfrak{e}_{6(2)}$ is done by further dimensional
reduction from $d=6$ to $d=4$. Projecting spinors is done in a
similar way and results in breaking $R$-symmetry algebra to
$\mathfrak{u}\left(1\right)$. The resulting   5-graded decomposition
of $\mathfrak{e}_{6(2)}$ is:
\begin{equation}
   \mathfrak{e}_{6(2)} = \tilde{\mathbf{1}} \oplus \tilde{\mathbf{20}} \oplus \left[
\mathfrak{su}(3,3) \oplus \mathfrak{so}(1,1) \right] \oplus \mathbf{20} \oplus \mathbf{1}
\end{equation}
\begin{equation}
\begin{split}
  \mathfrak{e}_{6(2)} &= \mathbf{1} \oplus
    \left(\begin{aligned} \left(\mathbf{2}, \mathbf{6}_v \right)\mspace{10mu} \\
                          \left(\mathbf{1}, \mathbf{4}_c \right)^+ \\
                          \left(\mathbf{1}, \mathbf{4}_s \right)^-
          \end{aligned}\right) \oplus \\
     &\left[ \Delta \oplus \left[
         \begin{aligned}
          &\mspace{20mu} \left( \mathbf{2}, \mathbf{4}_s\right)^+ \oplus   \left( \mathbf{2}, \mathbf{4}_c\right)^- \cr
          &\mathfrak{sp}\left(2, \mathbb{R}\right) \oplus \mathfrak{u}\left(1\right) \oplus \mathfrak{so}\left(4,2\right)
      \end{aligned} \right]  \right] \oplus
    \left(\begin{aligned} \left(\mathbf{2}, \mathbf{6}_v \right)\mspace{10mu} \\
                          \left(\mathbf{1}, \mathbf{4}_c \right)^+ \\
                          \left(\mathbf{1}, \mathbf{4}_s \right)^-
          \end{aligned}\right) \oplus
    \mathbf{1}
\end{split}
\end{equation}
where $+$ and $-$ refer to $\pm 1$ charges of
$\mathfrak{u}\left(1\right)$. Let  $\zeta^{\alpha}$ be a chiral
spinor of $\mathfrak{so}\left(4,2\right)$ and $\zeta^{\Dot{\alpha}}$
the corresponding  anti-chiral spinor. We shall combine these two
chiral spinors into one Majorana spinor $\psi^{A}$ of
$\mathfrak{so}\left(4,2\right)$. The decomposition of the The
generators of $\mathfrak{su}(3,3)$ in  $\mathfrak{g}^{0}$ subspace
read as follows
\begin{equation*}
\begin{split}
  M_{\mu\nu} &= 
            \eta_{\mu\rho} X^{\rho,a} \frac{\partial}{\partial X^{\nu,a}} -
                \eta_{\nu\rho} X^{\rho,a} \frac{\partial}{\partial X^{\mu,a}} -
             \psi^A {\left(\Gamma_{\mu\nu}\right)^A}_B \frac{\partial}{\partial \psi^B}\\
  H &= \zeta^{\alpha} \frac{\partial}{\partial \zeta^{\alpha}} -
            \zeta^{\Dot{\alpha}} \frac{\partial}{\partial \zeta^{\Dot{\alpha}}}
       = \psi^A {\left(\Gamma_7\right)^B}_A \frac{\partial}{\partial \psi^B} \\
  Q_{a, A} &= \epsilon_{ab} X^{\mu, a} {\left(\Gamma_\mu \right)^B}_A \frac{\partial}{\partial \psi^B} +
             \eta^{\mu\nu} \psi^B \left( C \Gamma_\mu \Gamma_7 \right)_{BA} \frac{\partial}{\partial X^{\nu, a}}
\end{split}
\end{equation*}
while $J_{ab}$ is defined as before and the charge conjugation matrix
is now symmetric $C^t = C$. These generators of
$\mathfrak{su}\left(3,3\right)$ satisfy the commutation relations
\begin{equation}
\begin{split}
  \left[ Q_{a,A}, Q_{b, B} \right] &= \frac{3}{2} \epsilon_{ab} C_{AB} H -
           \epsilon_{ab} \left(C\Gamma_{\mu\nu}\right)_{AB} M^{\mu\nu} -
                      \left(C\Gamma_7\right)_{AB} J_{ab} \\
  \left[ H, Q_{a,A} \right] &= {\left(\Gamma_7\right)^B}_A Q_{a, B} \\
\end{split}
\end{equation}
The chiral components of the generators of $Q_{a, A}$ are given by
\begin{equation*}
\begin{split}
Q_{a, \alpha} &= \epsilon_{ab} X^{\mu, b} {{\left( \Gamma_\mu
\right)}^{\Dot{\alpha}}}_{\alpha}
                \frac{\partial}{\partial \zeta^{\Dot{\alpha}}} +
                 \eta^{\mu\nu} \zeta^{\beta} \left( C \Gamma_\mu \right)_{\beta\alpha} \frac{\partial}{\partial X^{\nu, a}}\\
  Q_{a, \Dot{\alpha}} &= \epsilon_{ab} X^{\mu, b} {\left( \Gamma_\mu \right)^{\alpha}}_{\Dot{\alpha}}
                \frac{\partial}{\partial \zeta^{\alpha}} -
          \eta^{\mu\nu} \zeta^{\Dot{\beta}} \left( C \Gamma_\mu \right)_{\Dot{\beta}\Dot{\alpha}}
             \frac{\partial}{\partial X^{\nu, a}} \\
\end{split}
\end{equation*}
The  4-th order invariant of $\mathfrak{su}\left(3,3\right)$ written
in terms of $X$ and $\psi$ reads as follows:
\begin{equation}
\begin{split}
  \mathcal{I}_4 &= \eta_{\mu\nu} \eta_{\rho\tau} \epsilon_{ac}  \epsilon_{bd} X^{\mu, a} X^{\nu, b} X^{\rho, c} X^{\tau, d}
       - 2 X^{\mu, a} \epsilon_{ab} X^{\nu, b} \psi^A \psi^B \left( C \Gamma_{\mu\nu}\right)_{AB}
     \\  &+ \frac{1}{3} \eta^{\mu\nu}\eta^{\rho\tau} \psi^A \psi^B \left( C \Gamma_{\mu\rho}\right)_{AB}
        \psi^E \psi^F \left( C \Gamma_{\nu\tau}\right)_{EF}
\end{split}
\end{equation}
The spinorial generators of $\mathrm{E}_{6(2)}$ belonging to
$\mathfrak{g}^{-1}$ subspace  are realized as
\begin{equation}
   U_{A} = \frac{\partial}{\partial \psi^A} + \left(C \Gamma_7\right)_{AB} \psi^B \frac{\partial}{\partial y}
\end{equation}
which commute into the grade $-2$ generator $K_-$
\begin{equation}
   \left[ U_A, U_B \right] = 2 \left( C \Gamma_7\right)_{AB} K_-
\end{equation}
The commutators of the generators $Q_{a, A}$ belonging to grade zero
subspace with those in grade -1 subspace read as
\begin{equation}
\begin{split}
   \left[ Q_{a, A}, U_B \right] &= - \left(C \Gamma_\mu \Gamma_7 \right)_{AB} \eta^{\mu\nu} U_{\nu, a}\\
   \left[ Q_{a, A}, U_{\mu,b} \right] &=  - \epsilon_{ab} U_{B} {\left(\Gamma_\mu\right)^B}_A
\end{split}
\end{equation}
Commutation relations of the form $\left[ \mathfrak{g}^{-1},
\mathfrak{g}^{+1}\right] \subset \mathfrak{g}^0$ are
\begin{equation}
\begin{split}
   \left[ U_A, \Tilde{U}_B \right] &= -\frac{3}{2} \left(C \Gamma_7 \right)_{AB} H - C_{AB} \Delta +
           \left( C \Gamma_{\mu\nu}\right)_{AB} M^{\mu\nu} \\
   \left[ U_A, \Tilde{U}_{\mu, a}\right] &= {\left( C \Gamma_\mu C^{-1}\right)_A}^B Q_{a, B} \\
   \left[ \Tilde{U}_A, U_{\mu, a}\right] &= -{\left( C \Gamma_\mu C^{-1}\right)_A}^B Q_{a, B}
\end{split}
\end{equation}
Explicit expressions for positive grade generators are as follows:
\begin{equation}
\begin{split}
   \Tilde{U}_{\mu, a} &= -\frac{1}{4} \frac{\partial \mathcal{I}_4}{\partial X^{\mu, a}} \frac{\partial}{\partial y} -
    \eta_{\mu\nu}\epsilon_{ab} X^{\nu, b} y \frac{\partial}{\partial y} + y \frac{\partial}{\partial X^{\mu,a}}
    \\ &-\frac{1}{4} \frac{\partial^2 \mathcal{I}_4}{\partial X^{\mu, a}X^{\nu,b}} \eta^{\nu\rho}\epsilon^{bc}
    \frac{\partial}{\partial X^{\rho,c}} -
    \frac{1}{4} \frac{\partial^2\mathcal{I}_4}{\partial X^{\mu,a} \psi^A}
        \left(  \Gamma_7 C^{-1}\right)^{AB} \frac{\partial}{\partial \psi^B} \\ &-
     \eta_{\mu\nu}\epsilon_{ab} X^{\nu,b} \left( X^{\lambda,c}\frac{\partial}{\partial X^{\lambda,c}} +
         \psi^A \frac{\partial}{\partial \psi^A} \right)
\end{split}
\end{equation}
\begin{equation}
\begin{split}
   \Tilde{U}_{A} &= -\frac{1}{4} \frac{\partial \mathcal{I}_4}{\partial \psi^A} \frac{\partial}{\partial y} -
    \left( C \Gamma_7\right)_{AB} \psi^B y \frac{\partial}{\partial y} + y \frac{\partial}{\partial \psi^A}
    \\ &-\frac{1}{4} \frac{\partial^2 \mathcal{I}_4}{\partial \psi^A \psi^B} \left(\Gamma_7 C^{-1} \right)^{BC}
    \frac{\partial}{\partial \psi^C} -
    \frac{1}{4} \frac{\partial^2\mathcal{I}_4}{\partial X^{\mu,a} \psi^A}
        \eta^{\mu\nu} \epsilon^{ab} \frac{\partial}{\partial X^{\nu,b}} \\ &-
     \left( C \Gamma_7\right)_{AB} \psi^B \left( X^{\lambda,c}\frac{\partial}{\partial X^{\lambda,c}} +
         \psi^C \frac{\partial}{\partial \psi^C} \right)
\end{split}
\end{equation}
\begin{equation}
\begin{split}
  K_+ &= \frac{1}{2}\left( 2 y^2 - \mathcal{I}_4\right) \frac{\partial}{\partial y} +
          y \left( X^{\mu, a} \frac{\partial}{\partial X^{\mu,a}} + \psi^A \frac{\partial}{\partial \psi^A} \right) \\
     &- \frac{1}{4} \frac{\partial \mathcal{I}_4 }{\partial X^{\mu,a}} \epsilon^{ab} \eta^{\mu\nu}
      \frac{\partial}{ \partial X^{\nu,b}} -
    \frac{1}{4} \frac{\partial \mathcal{I}_4 }{\partial \psi^A} \left( \Gamma_7 C^{-1}\right)^{AB} \frac{\partial}{\partial \psi^B}
\end{split}
\end{equation}

\subsection{Geometric realization of the quasiconformal group
$F_{4(4)}$ }

 Further truncation to $\mathfrak{f}_{4(4)}$ is performed by reducing
from $d=4$ to $d=3$. The 5-grading in this case is
\begin{equation}
 \mathfrak{f}_{4(4)} =
   \tilde{\mathbf{1}} \oplus \tilde{\mathbf{14}} \oplus [
 \mathfrak{sp}(6,\mathbb{R}) \oplus \mathfrak{so}(1,1) ] \oplus \mathbf{14}
\oplus \mathbf{1 }
\end{equation}
\begin{equation}
\begin{split}
  \mathfrak{f}_{4(4)} &= \mathbf{1} \oplus
    \left(\begin{aligned} \left(\mathbf{2},  \mathbf{5} \right) \\
                          \left(\mathbf{1}, \mathbf{4} \right)
          \end{aligned}\right) \oplus \\
     &\left[ \Delta \oplus \left[
         \begin{aligned} & \mspace{35mu}\left( \mathbf{2},  \mathbf{4}\right) \cr
          \mathfrak{sp}&\left(2, \mathbb{R}\right) \oplus \mathfrak{so}\left(3,2\right)
 \end{aligned} \right]  \right] \oplus
    \left(\begin{aligned} \left(\mathbf{2}, \mathbf{5} \right) \\
                          \left(\mathbf{1}, \mathbf{4} \right)
      \end{aligned}\right) \oplus
    \mathbf{1}
\end{split}
\end{equation}
We use the same notations for spinors as above, assuming that now $A=1,\ldots,4$.
\begin{equation*}
\begin{split}
  M_{\mu\nu} &=
            \eta_{\mu\rho} X^{\rho,a} \frac{\partial}{\partial X^{\nu,a}} -
                \eta_{\nu\rho} X^{\rho,a} \frac{\partial}{\partial X^{\mu,a}} -
             \psi^A {\left(\Gamma_{\mu\nu}\right)^A}_B \frac{\partial}{\partial \psi^B}\\
  Q_{a, A} &= \epsilon_{ab} X^{\mu, a} {\left(\Gamma_\mu \right)^B}_A \frac{\partial}{\partial \psi^B} +
             \eta^{\mu\nu} \psi^B \left( C \Gamma_\mu \right)_{BA} \frac{\partial}{\partial X^{\nu, a}} \\
\end{split}
\end{equation*}
where $C^t = -C$.  The generators $Q_{a,A}$ close into
$\mathfrak{sp}\left(2,\mathbb{R}\right)\oplus
\mathfrak{so}\left(3,2\right)$ as fol\-lows:
\begin{equation}
   \left[ Q_{a,A}, Q_{b,B} \right] = \epsilon_{ab} \left(C\Gamma_{\mu\nu}\right)_{AB} M^{\mu\nu} + C_{AB} J_{ab}
\end{equation}
and transform under $\mathfrak{sp}\left(2,\mathbb{R}\right)\oplus
\mathfrak{so}\left(3,2\right)$ in the  $\left(\mathbf{2},
\mathbf{4}\right)$ representation. The generators
$Q_{a,A},M^{\mu\nu} $ and $J_{ab}$ form the
$\mathfrak{sp}(6,\mathbb{R})$ subalgebra.

Generators $K_- \in \mathfrak{g}^{-2}$ and $U_{\mu,a}\in
\mathfrak{g}^{-1}$  are as above and spinorial generators of
$\mathfrak{g}^{-1}$ are given by
\begin{equation}
   U_{A} = \frac{\partial}{\partial \psi^A} + \left(C\right)_{AB} \psi^B \frac{\partial}{\partial y}
\end{equation}
Spinorial generators form an Heisenberg subalgebra with charge
conjugation matrix $C$ serving as symplectic metric:
\begin{equation}
\left[U_{A}, U_{B} \right] = -2 C_{AB} K_-
\end{equation}
The generators $Q$ act on $\mathfrak{g}^{-1}$ subspace  as follows
\begin{equation}
\begin{split}
   \left[Q_{a,A}, U_{B} \right] &= \left(C\Gamma_\mu\right)_{AB} \eta^{\mu\nu} U_{\nu, a}\\
   \left[Q_{a,A}, U_{\mu, b} \right] &= -\epsilon_{ab} {\left(\Gamma_{\mu}\right)^B}_A U_{B}
\end{split}
\end{equation}
Quartic invariant $\mathcal{I}_4$ of $\mathfrak{sp}(6,\mathbb{R})$
in  $\mathfrak{sp}\left(2,\mathbb{R}\right)\oplus
\mathfrak{so}\left(3,2\right)$ basis is given by
\begin{equation}
 \mathcal{I}_4 =
   \eta_{\mu\nu} \eta_{\rho\tau} \epsilon_{ac}  \epsilon_{bd} X^{\mu, a} X^{\nu, b} X^{\rho, c} X^{\tau, d}
       - 2 X^{\mu, a} \epsilon_{ab} X^{\nu, b} \psi^A \psi^B \left( C \Gamma_{\mu\nu}\right)_{AB}
\end{equation}
Notice that the quartic term involving purely spinorial coordinates
,present in previous cases, now vanishes, since there is no
symmetric rank 4 invariant tensor of $\mathfrak{so}\left(3,2\right)
\simeq \mathfrak{sp}\left(4,\mathbb{R}\right)$ over its spinorial
representation space. Then the positive grade generators are
\begin{equation}
\begin{split}
     K_+ &= \frac{1}{2}\left( 2 y^2 - \mathcal{I}_4\right) \frac{\partial}{\partial y} +
          y \left( X^{\mu, a} \frac{\partial}{\partial X^{\mu,a}} + \psi^A \frac{\partial}{\partial \psi^A} \right) \\
     &- \frac{1}{4} \frac{\partial \mathcal{I}_4 }{\partial X^{\mu,a}} \epsilon^{ab} \eta^{\mu\nu}
      \frac{\partial}{ \partial X^{\nu,b}} +
    \frac{1}{4} \frac{\partial \mathcal{I}_4 }{\partial \psi^A}
       \left( C^{-1}\right)^{AB} \frac{\partial}{\partial \psi^B}
\end{split}
\end{equation}
\begin{equation}
\begin{split}
   \Tilde{U}_{\mu, a} &= -\frac{1}{4} \frac{\partial \mathcal{I}_4}{\partial X^{\mu, a}} \frac{\partial}{\partial y} -
    \eta_{\mu\nu}\epsilon_{ab} X^{\nu, b} y \frac{\partial}{\partial y} + y \frac{\partial}{\partial X^{\mu,a}}
    \\ &-\frac{1}{4} \frac{\partial^2 \mathcal{I}_4}{\partial X^{\mu, a}X^{\nu,b}} \eta^{\nu\rho}\epsilon^{bc}
    \frac{\partial}{\partial X^{\rho,c}} +
    \frac{1}{4} \frac{\partial^2\mathcal{I}_4}{\partial X^{\mu,a} \psi^A}
        \left( C^{-1}\right)^{AB} \frac{\partial}{\partial \psi^B} \\ &-
     \eta_{\mu\nu}\epsilon_{ab} X^{\nu,b} \left( X^{\lambda,c}\frac{\partial}{\partial X^{\lambda,c}} +
         \psi^A \frac{\partial}{\partial \psi^A} \right)
\end{split}
\end{equation}
\begin{equation}
\begin{split}
   \Tilde{U}_{A} &= -\frac{1}{4} \frac{\partial \mathcal{I}_4}{\partial \psi^A} \frac{\partial}{\partial y} +
     C_{AB} \psi^B y \frac{\partial}{\partial y} + y \frac{\partial}{\partial \psi^A}
    \\ &+\frac{1}{4} \frac{\partial^2 \mathcal{I}_4}{\partial \psi^A \psi^B} \left( C^{-1} \right)^{BC}
    \frac{\partial}{\partial \psi^C} -
    \frac{1}{4} \frac{\partial^2\mathcal{I}_4}{\partial X^{\mu,a} \psi^A}
        \eta^{\mu\nu} \epsilon^{ab} \frac{\partial}{\partial X^{\nu,b}} \\ &+
     \left( C\right)_{AB} \psi^B \left( X^{\lambda,c}\frac{\partial}{\partial X^{\lambda,c}} +
         \psi^C \frac{\partial}{\partial \psi^C} \right)
\end{split}
\end{equation}
Commutation relations of generators belonging to $\mathfrak{g}^{-1}$
and to $\mathfrak{g}^{+1}$ are
\begin{equation}
\begin{split}
  \left[ U_A, \Tilde{U}_B \right] &= \left(C \Gamma_{\mu\nu} \right)_{AB} M^{\mu\nu} - C_{AB} \Delta \\
  \left[ U_A, \Tilde{U}_{\mu,a} \right] &= - {\left( C \Gamma_\mu C^{-1} \right)_A}^B Q_{a, B} \\
  \left[ \Tilde{U}_A, U_{\mu,a} \right] &= {\left( C \Gamma_\mu C^{-1} \right)_A}^B Q_{a, B}
\end{split}
\end{equation}

\section{Minimal unitary realizations of the quasiconformal
groups $\mathrm{SO}(d+2,4)$}

In our earlier work \cite{GP} we constructed  the minimal unitary
representations of the exceptional groups discussed in the previous
section. In this section
 we will construct the minimal unitary representations of the groups
$SO\left(d+2, 4\right)$, corresponding to the quantization of their
geometric realizations as  quasiconformal groups given in section 5
following methods of \cite{GKN2,GP}. Let $X^\mu$ and $P_\mu$ be
canonical coordinates and momenta in
$\mathbb{R}^{\left(2,d\right)}$:
\begin{equation}
    \left[ X^\mu, P_\nu\right] = i {\delta^\mu_\nu}
\end{equation}
In the notation of section 5 we identify $X^{\mu, a=1}$ to be
coordinates $X^\mu$, and $P_\mu = \eta_{\mu\nu} X^{\nu, a=2}$ to be
conjugate momenta. Also let $x$ be an additional ``cocycle''
coordinate and $p$ be its conjugate momentum:
\begin{equation}
   \left[ x, p \right] = i
\end{equation}
The grade zero generators ( $M_{\mu\nu}, J_{\pm}, J_0$), grade $-1$
generators ( $ U_{\mu} , V^{\mu}$), grade $-2$ generator $K_-$ and
the 4-th order  invariant $\mathcal{I}_4 $ of the semisimple part of
the grade zero subalgebra
 are realized as follows:
\begin{equation}
\begin{split}
  & \begin{aligned}
      M_{\mu\nu} &= i \eta_{\mu\rho} X^\rho P_\nu - i \eta_{\nu\rho} X^\rho P_\mu  \\
      U_\mu &= x P_\mu  \qquad V^\mu = x X^\mu \\
      K_- & = \frac{1}{2} x^2
   \end{aligned}
\qquad\qquad
    \begin{aligned}
        J_0 &= \frac{1}{2} \left( X^\mu P_\mu + P_\mu X^\mu \right) \\
        J_- &= X^\mu X^\nu \eta_{\mu\nu} \\
        J_+ &= P_\mu P_\nu \eta^{\mu\nu}
    \end{aligned} \\
   &\phantom{aga} \mathcal{I}_4 = \left(X^\mu X^\nu \eta_{\mu\nu}\right) \left(P_\mu P_\nu \eta^{\mu\nu}\right) +
    \left(P_\mu P_\nu \eta^{\mu\nu}\right) \left(X^\mu X^\nu \eta_{\mu\nu}\right) \\
     & \phantom{aga also} -  \left(X^\mu P_\mu\right)\left( P_\nu X^\nu \right) -
       \left(P_\mu X^\mu\right) \left( X^\nu P_\nu\right)
\end{split}
\end{equation}
Using the quartic invariant we define the grade +2 generator as
\begin{equation}
    K_+ = \frac{1}{2} p^2 + \frac{1}{4 \, y^2} \left( \mathcal{I}_4 + \frac{d^2+3}{2} \right)
\end{equation}
It is easy to verify that the generators $M_{\mu\nu}$ and
$J_{0,\pm}$ satisfy the commutation relations of
$\mathfrak{so}\left(d,2\right)\oplus \mathfrak{sp}\left(2,
\mathbb{R}\right)$
\begin{equation}
\begin{split}
       \left[ M_{\mu\nu}, M_{\rho\tau} \right] &= \eta_{\nu\rho} M_{\mu\tau} - \eta_{\mu\rho} M_{\nu\tau} + \eta_{\mu\tau} M_{\nu\rho} - \eta_{\nu\tau} M_{\mu\rho}   \\
     \left[ J_0, J_\pm \right] &= \pm 2 i J_\pm \qquad
          \left[ J_-, J_+ \right] = 4 i J_0
\end{split}
\end{equation}
under which coordinates $X^\mu$ and momenta $P^\mu$ transform as
Lorentz vectors and form doublets of the symplectic group
$Sp(2,\mathbb{R}$) :
\begin{equation}
     \begin{aligned}
          \left[ J_0, V^\mu \right] &=  - i V^\mu \\
          \left[ J_0, U_\mu \right] &=  + i U_\mu
     \end{aligned}
         \quad
     \begin{aligned}
         \left[ J_-, V^\mu \right] &= 0 \\
         \left[ J_-, U_\mu \right] &= 2 i \eta_{\mu\nu} V^\nu
     \end{aligned}
          \quad
     \begin{aligned}
         \left[ J_+, V^\mu \right] &= -2 i \eta^{\mu\nu} U_\nu \\
         \left[ J_+, U_\mu \right] &= 0
     \end{aligned}
\end{equation}
The generators in the subspace $\mathfrak{g}^{-1}\oplus
\mathfrak{g}^{-2}$ form a Heisenberg algebra
\begin{equation}
    \left[ V^\mu, U_\nu \right] = 2 i {\delta^\mu}_\nu \,  K_- \,.
\end{equation}
Define the  grade $+1$ generators  as
\begin{equation}
   \Tilde{V}^\mu = -i \left[ V^\mu, K_+ \right]  \qquad
   \Tilde{U}_\mu = -i \left[ U_\mu, K_+ \right]
\end{equation}
which explicitly read as follows
\begin{equation}
\begin{split}
    \Tilde{V}^\mu &= p X^\mu + \frac{1}{2} x^{-1}  \left(  P_\nu X^\lambda X^\rho  +
     X^\lambda X^\rho  P_\nu \right) \eta^{\mu\nu} \eta_{\lambda\rho} \\
       &- \frac{1}{4} x^{-1} \left( X^\mu \left(X^\nu P_\nu + P_\nu X^\nu\right) +
      \left(X^\nu P_\nu + P_\nu X^\nu\right) X^\mu\right) \\
    \Tilde{U}_\mu &= p P_\mu - \frac{1}{2} x^{-1}
   \left(  X^\nu P_\lambda P_\rho  +  P_\lambda P_\rho  X^\nu \right) \eta_{\mu\nu} \eta^{\lambda\rho} \\
       &+ \frac{1}{4} x^{-1} \left( P_\mu \left(X^\nu P_\nu + P_\nu X^\nu\right) +
      \left(X^\nu P_\nu + P_\nu X^\nu\right) P_\mu\right) .
\end{split}
\end{equation}
Then one finds that the generators in $\mathfrak{g}^{+1} \oplus
\mathfrak{g}^{+2} $ subspace form an isomorphic Heisenberg algebra
\begin{equation}
  \left[ \Tilde{V}^\mu, \Tilde{U}_\nu \right] = 2 i {\delta^\mu}_\nu K_+ \qquad
 V^\mu = i \left[ \Tilde{V}^\mu, K_- \right] \qquad  U_\mu = i \left[ \Tilde{U}_\mu, K_- \right] .
\end{equation}
Commutators $\left[ \mathfrak{g}^{-1}, \mathfrak{g}^{+1} \right]$
close into $\mathfrak{g}^0$ as follows
\begin{equation}
\begin{split}
   \left[ U_\mu, \Tilde{U}_\nu \right] &= i \eta_{\mu\nu} J_- \qquad
   \left[ V^\mu, \Tilde{V}^\nu \right] = i \eta^{\mu\nu} J_+ \\
   \left[ V^\mu, \Tilde{U}_\nu \right] &= 2 \eta^{\mu\rho} M_{\rho\nu} + i {\delta^\mu}_\nu \left( J_0 +  \Delta \right)
   \\
    \left[ U_\mu, \Tilde{V}^\nu \right] &= - 2 \eta^{\nu\rho} M_{\mu\rho} + i {\delta^\nu}_\mu \left( J_0 -  \Delta \right)
\end{split}
\end{equation}
where $\Delta$ is the  generator that determines the 5-grading
\begin{equation}
   \Delta = \frac{1}{2} \left( x p + p x\right)
\end{equation}
such that
\begin{equation}
   \left[ K_-, K_+ \right] = i \Delta \qquad \left[ \Delta, K_\pm \right] = \pm 2 i K_\pm \quad
\end{equation}
\begin{equation}
 \left[ \Delta, U_\mu \right] = - i U_\mu \quad \left[ \Delta, V^\mu \right] = - i V^\mu  \quad
 \left[ \Delta, \Tilde{U}_\mu \right] =  i \Tilde{U}_\mu \quad
 \left[ \Delta, \Tilde{V}^\mu \right] =  i \Tilde{V}^\mu  \quad
\end{equation}
The quadratic Casimir operators of subalgebras
$\mathfrak{so}\left(d,2\right)$, $\mathfrak{sp}\left(2,
\mathbb{R}\right)_J$ of grade zero subspace and
$\mathfrak{sp}\left(2, \mathbb{R}\right)_K$ generated by $K_{\pm}$
and $\Delta$ are
\begin{equation}
 \begin{aligned}
   M_{\mu\nu} M^{\mu\nu} &= - \mathcal{I}_4 - 2 \left(d+2\right) \cr
   J_- J_+ +  J_+ J_- - 2  \left(J_0\right)^2 &= \mathcal{I}_4 + \frac{1}{2}\left(d+2\right)^2 \cr
    K_- K_+ + K_+ K_- - \frac{1}{2} \Delta^2 &= \frac{1}{4} \mathcal{I}_4 + \frac{1}{8} \left(d+2\right)^2
 \end{aligned}
\end{equation}
Note that they all reduce to $\mathcal{I}_4$ modulo some additive
constants. Noting also that
\begin{equation}
  \left( U_\mu \Tilde{V}^\mu + \Tilde{V}^\mu U_\mu  - V^\mu \Tilde{U}_\mu - \Tilde{U}_\mu V^\mu \right) =
     2 \mathcal{I}_4 + \left(d+2\right)\left(d+6\right)
\end{equation}
we conclude that there exists a family of degree 2 polynomials in
the enveloping algebra of $\mathfrak{so}\left(d+2,4\right)$ that
degenerate to a c-number for the  minimal unitary  realization, in
accordance with Joseph's theorem \cite{Joseph}:
\begin{equation}\label{eq:JosephIdeal}
\begin{split}
   M_{\mu\nu} M^{\mu\nu} &+ \kappa_1 \left( J_- J_+ +  J_+ J_- - 2  \left(J_0\right)^2 \right)
   +  4 \kappa_2 \left(K_- K_+ + K_+ K_- - \frac{1}{2} \Delta^2\right) \\&-
     \frac{1}{2}\left(\kappa_1+\kappa_2-1\right)
    \left( U_\mu \Tilde{V}^\mu + \Tilde{V}^\mu U_\mu  - V^\mu \Tilde{U}_\mu - \Tilde{U}_\mu V^\mu \right)
  \\ &= \frac{1}{2}\left(d+2\right)\left( d + 2 - 4\left(\kappa_1+\kappa_2\right) \right)
\end{split}
\end{equation}

The quadratic Casimir of $\mathfrak{so}\left(d+2, 4\right)$
corresponds to the choice $2 \kappa_1 = 2 \kappa_2 = -1$ in \eqref{eq:JosephIdeal}. Hence
the eigenvalue of the quadratic Casimir for the minimal unitary
representation is equal to $\frac{1}{2} \left(d+2\right) \left(d+6\right)$.

{\bf Acknowledgement:} This work was supported in part by the National
Science Foundation under grant number PHY-0245337. Any opinions,
findings and conclusions or recommendations expressed in this material
are those of the authors and do not necessarily reflect the views of
the National Science Foundation.


\begin{thebibliography}{99}
\bibitem{mg75} M.~G\"unaydin, ``\emph{Exceptional realizations of Lorentz group: supersymmetries and leptons},''
     Nuovo Cim.\ {\bf 29A} (1975) 467.

\bibitem{mg80}  M. G\"unaydin,''\emph{Quadratic Jordan formulation of quantum mechanics and construction of Lie
  (super)algebras from Jordan (super)algebras},'' Ann. Israel Physical Society {\bf 3} (1980) 279.

\bibitem{mg91}  M. G\"unaydin, "The Exceptional Superspace and the Quadratic Jordan Formulation of
    Quantum Mechanics",  in ``Elementary Particles and the Universe:
     {\it Essays in Honor of Murray Gell-Mann},'' ed. by J.H. Schwarz
    (Cambridge University Press,  1991); M.~G\"unaydin,
     ``On An Exceptional Nonassociative Superspace,'' J.\ Math.\ Phys.\  {\bf 31}, 1776 (1990).
\bibitem{mg92} M. G\"{u}naydin,
      ``\emph{Generalized conformal and superconformal group actions and Jordan algebras},''
    Mod.\ Phys.\ Lett.\  {\bf A8}, 1407 (1993) [hep-th/9301050].


\bibitem{Mack69}
G.~Mack, A.~Salam.
\newblock Finite component field representations of the conformal group.
\newblock \emph{Ann. Phys.}, \textbf{53}, 174 (1969)
\bibitem{Mack77}
G.~Mack.
\newblock {All unitary ray representations of the conformal group SU(2,2) with
  positive energy.}
\newblock \emph{Commun. Math. Phys.}, \textbf{55}, 1 (1977)
and the references therein.
\bibitem{Guna99}
M.~G{\"u}naydin, ``\emph{AdS/CFT dualities and the unitary representations
of non-compact groups and supergroups: Wigner versus Dirac}'', invited
talk in the proceedings of the VIth International Wigner Symposium
(Istanbul, 1999), [hep-th/0005168], ed. by M. Arik,  Bogazici
University Press, 2002, pp. 55-69.

\bibitem{Koecher} M.~Koecher, ``\emph{On Lie algebras defined by Jordan algebras},'' Aarhus Univ. Lecture notes,
    Aarhus (1967);
\newblock {\"Uber eine Gruppe von rationalen Abbildungen}.
\newblock \emph{Invent. Math.}, \textbf{3}, 136 (1967)


\bibitem{GKN1}
  M.~G\"unaydin, K.~Koepsell and H.~Nicolai,
  ``\emph{Conformal and quasiconformal realizations of exceptional Lie groups},''
  Commun.\ Math.\ Phys.\  {\bf 221}, 57 (2001) [arXiv:hep-th/0008063].

\bibitem{GP}
  M.~G\"unaydin and O.~Pavlyk,
  ``\emph{Minimal unitary realizations of exceptional U-duality groups and their
  subgroups as quasiconformal groups},''
  JHEP {\bf 0501}, 019 (2005)
  [arXiv:hep-th/0409272].

\bibitem{li2000}
 Jian-Shu Li, \emph{Minimal representations and reductive dual pairs},
in \emph{Representation theory of Lie groups}, IAS/Park City
Mathematics Series Volume 8, J.~Adams and D.~Vogan eds., AMS
Publications, 2000.

\bibitem{PKW}
D.~Kazhdan, B.~Pioline and A.~Waldron, \emph{Minimal
representations, spherical vectors and exceptional theta series, I},
Comm. Math. Phys. {\bf 226}( 2002) 1 [hep-th{0107222}].
\bibitem{PW}
B.~Pioline and A.~Waldron, \emph{The automorphic membrane}, JHEP
{\bf 06}{2004}{009} [hep-th{0404018}].

  \bibitem{Freu54}
H.~Freudenthal.
\newblock {Beziehungen der ${E_7}$ und ${E_8}$ zur Oktavenebene. I.}
\newblock \emph{Nederl. Akad. Wet., Proc., Ser. A}, \textbf{57}, 218 (1954)
\bibitem{Meyb68}
K.~Meyberg.
\newblock {Eine Theorie der Freudenthalschen Tripelsysteme. I, II.}
\newblock \emph{Nederl. Akad. Wet., Proc., Ser. A}, \textbf{71}, 162 (1968)

\bibitem{KanSko82}
I.~Kantor, I.~Skopets.
\newblock {Some results on Freudenthal triple systems.}
\newblock \emph{Sel. Math. Sov.}, \textbf{2}, 293 (1982)

\bibitem{GST1} M.~G\"unaydin, G.~Sierra and P.~K.~Townsend,
   ``\emph{The geometry of N=2 Maxwell-Einstein Supergravity and Jordan algebras},''
   Nucl.\ Phys.\ B {\bf 242}, 244 (1984);\newblock ``Exceptional Supergravity Theories and the Magic
   Square.''
\newblock \emph{Phys. Lett.}, \textbf{B133}, 72 (1983);
\bibitem{Sierra} G.~ Sierra,
      ``\emph{An application of the theories of Jordan algebras and Freudenthal \
        triple systems to particle and strings},''
        Class.\ Quan.\ Grav. {\bf 4} (1987) 227.

\bibitem{ferrar} J.C. Ferrar, ``\emph{Strictly regular elements in
Freudenthal triple systems}'', Trans. Am. Math. Soc. {\bf 174} (1972)
313-331.
\bibitem{McCrimmon} K.~McCrimmon, "Norms and noncommutative Jordan algebras"
, Pacific Journal of Mathematics, {\bf 15} (1965) 925.

\bibitem{FrSprTits} H.~Freudenthal,"Beziehungen der $E_7$ und $E_8$ zur Oktavenebene. I."
 Nederl. Akad. Wetensch. Proc. Ser. {\bf 57A} (1954) 218;
 "Beziehungen der $E_7$ und $E_8$ zur Oktavenebene. VIII. ",\emph{i.b.i.d.},
 {\bf 62A} (1959) 447;
 T.~A.~Springer,," Characterization of a class of cubic forms",   \emph{i.b.i.d.} {\bf 65A} (1962) 259;
 J.~Tits, "Une classe d'Alg\`ebres de Lie en relation avec les alg\`ebres de
 Jordan",
\emph{i.b.i.d.} {\bf 65A} (1962) 530; "Alg\`ebres alternatives,
Alg\`ebres de Jordan et algebres de Lie
 exceptionelles", \emph{i.b.i.d.} {\bf 69A}
(1966) 223.
\bibitem{Faul71}
J.~Faulkner.
\newblock {A construction of Lie algebras from a class of ternary algebras}.
\newblock \emph{Trans. Am. Math. Soc.}, \textbf{155}, 397 (1971)

\bibitem{Schafers} R.~D.~Schafer,"Cubic forms permitting a new type of composition,"
 Journal of Mathematics and Mechanics, {\bf 10}, (1961);
  see also T.~A.~Springer,"On a class of Jordan algebras", Proc. Nederl. Akad. Wetensch. {\bf 62A} (1959) 254.

\bibitem{AllFau84}
B.~Allison, J.~Faulkner.
\newblock {A Cayley-Dickson process for a class of structurable algebras.}
\newblock \emph{Trans. Am. Math. Soc.}, \textbf{283}, 185 (1984)

\bibitem{Freu85}
H.~Freudenthal.
\newblock {Oktaven, Ausnahmegruppen und Oktavengeometrie.}
\newblock \emph{Geom. Dedicata}, \textbf{19}, 7 (1985)


\bibitem{four}
  B.~de Wit, P.~G.~Lauwers, R.~Philippe, S.~Q.~Su and A.~Van Proeyen,
  ``\emph{Gauge and matter fields coupled to $N$=2 supergravity},''
  Phys.\ Lett.\ B {\bf 134} (1984) 37.
\bibitem{MG2005} M.~Gunaydin,
  ``Unitary realizations of U-duality groups as conformal and quasiconformal
  groups and extremal black holes of supergravity theories,''
  arXiv:hep-th/0502235, invited talk in the Proceedings of the XIXth Max Born Symposium, AIP
  conference series No: 767, ed. by J. Lukierski and D. Sorokin ( 2005).
\bibitem{GKN2}
  M.~Gunaydin, K.~Koepsell and H.~Nicolai,
  ``The Minimal Unitary Representation of $E_{8(8)}$,''
  Adv.\ Theor.\ Math.\ Phys.\  {\bf 5}, 923 (2002)
  [arXiv:hep-th/0109005].
\bibitem{Joseph}
A.~Joseph.
\newblock {The minimal orbit in a simple Lie algebra and its associated maximal
  ideal.}
\newblock \emph{Ann. Sci. Ec. Norm. Super., IV. Ser.}, \textbf{9}, 1 (1976)

\end{thebibliography}
\end{document}